\documentclass[fleqn,usenatbib]{mnras}

\usepackage{newtxtext,newtxmath}

\usepackage[T1]{fontenc}
\usepackage{ae,aecompl}

\usepackage{graphicx}	
\usepackage{amsmath}	
\usepackage{amssymb}	

\usepackage[switch]{lineno}

\title[Absorption of $\gamma$-rays by the torus in FSRQs]{Probing the absorption of gamma-rays by IR radiation from the dusty torus in FSRQs with the Cherenkov Telescope Array}

\author[G. Galanti et al.]{
Giorgio Galanti,$^{1}$\thanks{E-mail: gam.galanti@gmail.com (GG)}
Marco Landoni,$^{1}$
Fabrizio Tavecchio$^{1}$
and Stefano Covino$^{1}$
\\
$^{1}$INAF, Osservatorio Astronomico di Brera, Via Emilio Bianchi 46, I -- 23807 Merate, Italy
}

\date{Accepted XXX. Received YYY; in original form ZZZ}

\pubyear{2019}

\begin{document}
\label{firstpage}
\pagerange{\pageref{firstpage}--\pageref{lastpage}}
\maketitle

\begin{abstract} 
Within the classical emission model, where the emission region is placed within the broad line region (BLR), flat spectrum radio quasars (FSRQs) were believed not to emit photons with energies above few tens of GeV because of the absorption with the optical-UV photons from the BLR. However, photons with observed energies up to about $300 \, \rm GeV$ have been detected for few FSRQs, whose most iconic example is PKS 1441+25 at redshift $z = 0.94$. The most conservative explanation for these observations is that the emission occurs at distances comparable to the size of the dusty torus.
In this case, absorption of high-energy gamma-ray photons for energies above $200-300 \, {\rm GeV}$ is dominated by the interaction with infrared radiation emitted by the torus. We investigate if current observational data about FSRQs in flaring state can give us information about: (i) the importance of the torus absorption and (ii) the properties of the torus i.e. its temperature and its geometry. We find that present data do not arrive at energies where the torus influence is prominent and as a result it is currently hardly possible to infer torus properties from observations. However, with dedicated simulations, we demonstrate that observations with the forthcoming Cherenkov Telescope Array (CTA) will be able to constrain the torus parameters (temperature and geometry).
\end{abstract}





\section{Introduction}

The study of relativistic jets is a fascinating subject that involves the comprehension of several basic astrophysical processes, such as particle acceleration, plasma physics, black hole dynamics (e.g.~\citealt{jet1,jet2}). Blazars are  extragalactic sources whose extreme phenomenology is attributed to a favorable geometry in which a relativistic jet produced by an active galactic nucleus is almost pointing toward the Earth. The relativistic beaming of the non-thermal emission produced in the jet is responsible for the large observed fluxes and the extremely rapid variability~\citep{BlandfordRees1978}. The powerful emission of these sources, spanning all the electromagnetic spectrum, from radio waves up to the very-high-energy (VHE) gamma-ray band can be used to effectively probe the jet and its physical processes.

The emission from blazars shows, when plotted in the $\nu F_{\nu}$ representation (the so-called spectral energy distribution, SED), two broad components (or ``bumps''), one peaking in the IR-UV band and the other one with the maximum at gamma-ray energies. Leptonic models attributes the two components to non-thermal synchrotron and inverse Compton (IC) emission from relativistic electrons in the jet. Hadronic scenarios assume the existence of a component of high-energy protons directly emitting through synchrotron or interacting with photons through photomeson reactions (e.g.~\citealt{Muecke2003}). Blazars are historically divided in two groups, defined by the properties of the optical spectra. Specifically, Flat Spectrum Radio Quasars (FSRQs) are powerful sources whose optical spectra display the intense broad emission lines characteristics of quasars. These lines are thought to be produced by clouds of gas orbiting around the central supermassive black hole (BH) in the so-called broad line region (BLR). The gas is photoionized by the intense UV continuum from the innermost regions of the accretion disk surrounding the BH and re-emits the radiation in typical emission lines Doppler-broadened by their rapid motion. BL Lac objects, on the other hand, are less powerful sources which display rather weak optical lines (sometimes they are even absent). This property is often attributed to the presence of a radiatively inefficient accretion flow (RIAF) onto the central BH (e.g.~\citealt{gmt09,righi19}). The emission from BL Lac can extend well above the TeV band and, indeed, these sources represent the majority of the extragalactic objects detected by the current generation of Imaging Atmospheric Cherenkov Telescope arrays (IACT), H.E.S.S., MAGIC and VERITAS.

The scarcity of detections of FSRQs in the VHE band can be attributed to a combination of several causes (e.g.~\citealt{tavecchio17}). Typically, FSRQs are characterized by redshifts larger than those of BL Lac objects, implying the severe reduction of the observed flux because of the absorption of VHE photons interacting via the pair production reaction $\gamma \gamma\to e^+e^-$ with the low energy radiation of the extragalactic background light (EBL; e.g.~\citealt{costamante13}). Moreover, the  intrinsic gamma-ray spectra of FSRQs  are in general softer than those of BL Lacs, as described by the so-called {\it blazar sequence}~\citep{fossati98,blazseq2}. The softening of the spectra is very pronounced at energies above few tens of GeV, fact explained with the on-set of the Klein-Nishina regime in the leptonic models (e.g.~\citealt{tg08}) coupled with the absorption of the gamma-ray photons interacting with the UV radiation field of the BLR (e.g.~\citealt{poutanen10}). The latter effect is important if the high-energy emission occurs within the BLR radius, commonly estimated to lie at distances of the order of $0.1-1 \, \rm pc$. Since most of the models accounting for the powerful high-energy emission of FSRQs assumed that the emission occurs well inside the BLR, it was commonly expected that the FSRQs cannot be bright above 100 GeV (e.g.~\citealt{tavecchio2010}). For this reason, the detection at VHE of a handful of FSRQs was a surprise. Although there are suggestions (e.g.~\citealt{PKS1510LUM}) supporting the idea that emission could occur at several parsecs from the core (at scales probed by VLBA observations), the most conservative option to avoid the absorption caused by the BLR is to place the emission region (at least for some sources and/or during particular states) just beyond the BLR, where the dominating ambient radiatiation is the infrared field provided by a torus of dust heated at about 1000 K by the disk radiation. This view has been recently confirmed by the analysis of {\it Fermi}/LAT spectra of a sample of bright FSRQ, which does not reveal the expected absorption at $10-20 \, \rm GeV$~\citep{costamante18}. We note that a dissipation region beyond the BLR is in agreement with numerical simulations of jet acceleration (e.g.~\citealt{komissarov2007}), which describe a relatively slow acceleration phase, typically ending at pc scale.
A back-on the envelope calculation shows that the low energy of the IR photons from the torus implies that the threshold energy for the pair-production reaction to occur increases at several hundreds of GeV. In fact, close to threshold, the typical photon energy of the (black body-like) IR radiation field interacting with the gamma-ray photons is $\langle \epsilon \rangle \simeq 2.82 \; kT \simeq 0.25 \; T_3$ eV (where the notation $Q=Q_X \, 10^X $ in cgs units has been introduced), implying that absorption becomes relevant for photons of energy $E=m_e^2c^4/\langle \epsilon \rangle =1 \; T_3^{-1}$ TeV. Therefore, if the gamma-ray photons are produced beyond the BLR but still inside the radiation field of the torus (i.e. at distances lower than few parsecs from the BH) we expect absorption to be visible in spectrum at (intrinsic) energies around 1 TeV.  Current detections do not allows us to reach such high energies. The best cases are those of the FSRQs PKS 1222+216 (at $z=0.432$) and PKS 1441+25 (at $z=0.940$) whose spectra detected by MAGIC during exceptional flaring states~\citep{PKS1222,PKS1441} extend to intrinsic energies below 600 GeV. One foresees to obtain much better results with the Cherenkov Telescope Array (CTA;~\citealt{CTA13,hofmann17}), the first observatory for the VHE gamma-ray astronomy presently under construction. In particular, the combination of the low energy threshold ensured by the Large Sized Telescopes (LSTs) and the superb sensitivity around 1 TeV provided by large number of Medium Size Telescopes (MSTs) will allow one to detect and study a large number of FSRQs with unprecedented accuracy. 

While simple theoretical models of the gamma-ray absorption caused by the torus radiation field are already available in literature (e.g.~\citealt{torus1}), an application to the currently available VHE spectra of the FSRQs and a study of the potentialities of the CTA is still lacking. With our work we would like to fill this gap. In particular, we will present a simple theoretical set-up to model the expected absorption by the IR radiation field and we will apply it to the best VHE spectra available for some FSRQs to investigate the constraints that can be derived on the main parameters describing the torus system. In the second part of the paper we will use the model spectra to simulate observations with the CTA and derive the constraints that can be inferred.  

The paper is organized as follows. In Sect. 2 we present our torus model, we calculate the optical depth due to absorption caused by torus IR radiation and we study its consequences on FSRQ spectra. In Sect. 3 we calculate the observed spectra of all considered FSRQs with different choices of the torus temperature. We calculate the observed spectrum energy bins for all FSRQs (PKS 1510-089, PKS 1222+216 and PKS 1441+25) and we perform a statistical analysis to understand CTA capability in determining torus parameters in Sect. 4, while we discuss our results and draw our conclusions in Sect. 5.

Throughout the paper, the following cosmological  parameters are assumed: $H_0=70$ km s$^{-1}$ Mpc$^{-1}$, $\Omega_{\rm M}=0.3$, $\Omega_{\Lambda}=0.7$.

\section{Optical depth}
In this section we closely follow the model and definitions developed in~\citet{torus1} for the torus design (see also references therein), while for the calculation of the torus optical depth we adapt to this specific case what is discussed in~\citet{dgrTransp}.

\begin{figure}       
\begin{center}
\includegraphics[width=.45\textwidth]{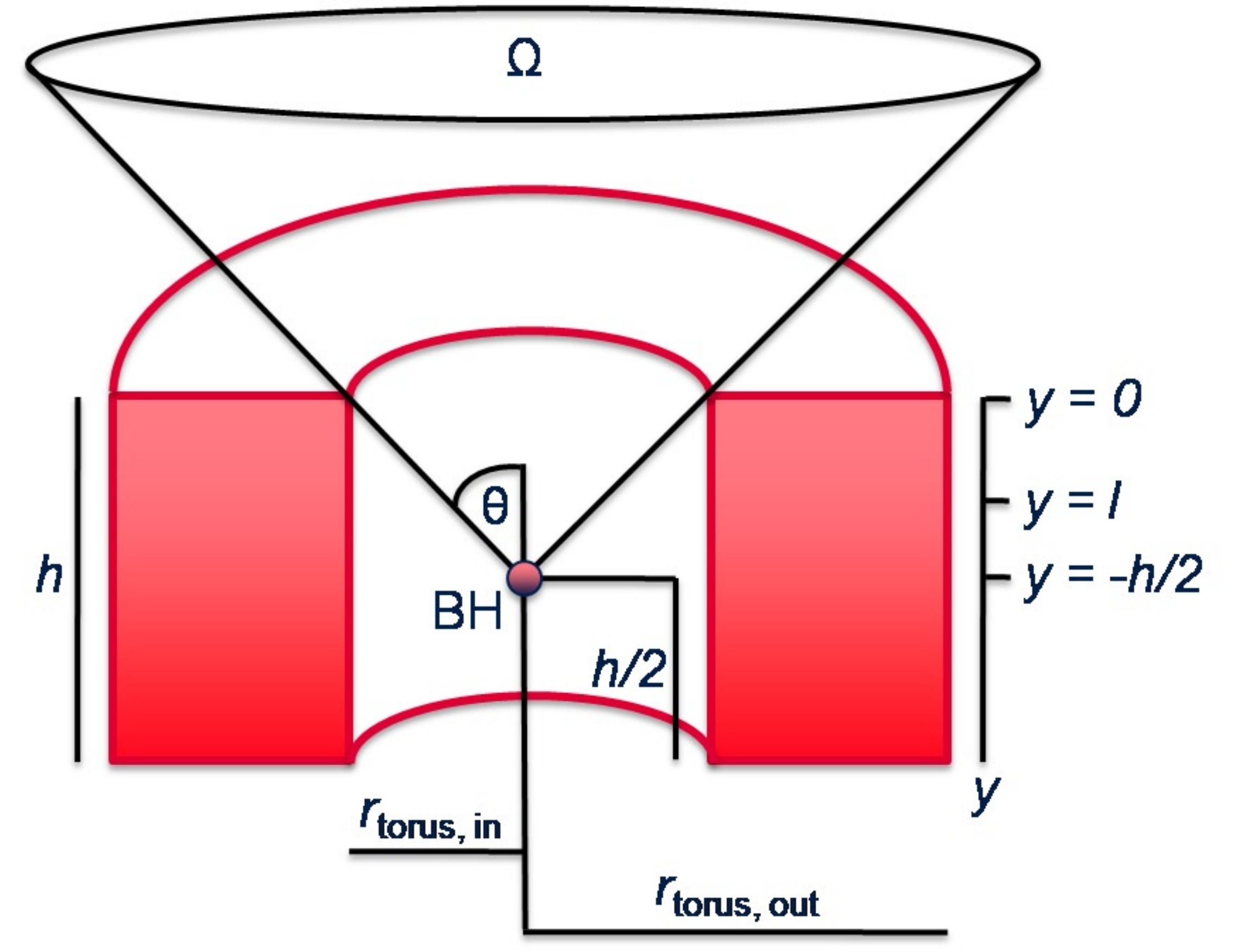}
\end{center}
\caption{\label{geom} 
Torus vertical section indicating the torus geometry. See the text for definition of different quantities.}
\end{figure}

The dust surrounding an AGN is well described as organized in a torus emitting as a black body in the infrared energy band. As a consequence, in first approximation the emission of the torus can be described by two parameters: the temperature $T$ and the covering factor $f_{\rm cov}$ which is related to the torus geometry as it will be clear in the following. We consider the torus as centered on the black hole and aligned to the jet axis. For simplicity, we approximate the torus with a rectangular cross section with inner radius $r_{\rm torus, in}$, outer radius $r_{\rm torus, out}$ and height $h$ as in Fig.~\ref{geom}.

The parameter $f_{\rm cov}$ measures how much of the disk radiation is covered by the torus and it is linked to the solid opening angle $\Omega$ (see Fig.~\ref{geom}) by the relation
\begin{equation}
\label{fcov}
f_{\rm cov}=1-\frac{\Omega}{4\pi}=\frac{2 \times 2\pi \int_{\theta}^{\pi/2}{\rm sin}\, \theta_1 \, {\rm d}\theta_1}{4\pi}={\rm cos}\,\theta~,
\end{equation}
which sets $f_{\rm cov}$ in the open interval $]0,1[$ while $\theta$ is the opening angle subtended by $\Omega$ (see Fig.~\ref{geom}). The upper and lower bounds of the interval are not physical. In fact, the infimum of the interval to which $f_{\rm cov}$ may belong corresponds to a solid opening angle $\Omega=4\pi$ which means a null torus thickness ($h=0$), while the supremum corresponds to a solid opening angle $\Omega=0$ which means an infinity torus thickness ($h \to \infty$). The inner radius of the torus $r_{\rm torus, in}$ depends on the disk luminosity $L_{\rm disk}$ and on the torus dust temperature $T$~\citep{rtorusRel}: accordingly it reads
\begin{equation}
\label{t1}
r_{\rm torus, in} \approx 0.316 \, T_{1500}^{-2.8} \, L_{\rm disk,45}^{1/2} \, {\rm pc}~,
\end{equation}
where $T_{1500}$ is the dust temperature in units of $1500 \, \rm K$ and $L_{\rm disk,45}$ is the disk luminosity in units of $10^{45} \, \rm erg \, s^{-1}$. We take $r_{\rm torus, out}=2 \, r_{\rm torus, in}$, while $h$ is directly linked to the value of $f_{\rm cov}$ i.e. to the opening solid angle $\Omega$. In fact, it is a simple exercise of trigonometry to get that
\begin{equation}
\label{t2}
h=2 \, r_{\rm torus, in} \, {\rm cot} \, \theta~,
\end{equation}
where
\begin{equation}
\label{t3}
{\rm cot} \, \theta=\frac{4\pi-\Omega}{\sqrt{8\pi\Omega-\Omega^2}}~.
\end{equation}
As discussed above, in our model we place the photon emission region beyond the BLR in order to avoid BLR absorption. Therefore, it is important to determine also the position of the BLR: in this fashion, we use a simple relation where the BLR radius $r_{\rm BLR}$ scales with the square root of $L_{\rm disk}$ as~\citep{gtRelLum}
\begin{equation}
\label{t4}
r_{\rm BLR} \approx 10^{17} \, L_{\rm disk,45}^{1/2} \, {\rm cm}~.
\end{equation}
The emission position $r_{\rm em}$ is a free parameter of the model. A complete exploration of the entire parameter space is beyond the aim of this paper.  For definiteness, we fix $r_{\rm em}=2\,r_{\rm BLR}$. The emission region may in principle be placed also beyond the zone characterized by a torus strong influence, as argued by e.g.~\citet{PKS1510LUM}. In this case, spectra would not be affected by intrinsic absorption.

We want to calculate the optical depth $\tau$ of a VHE photon produced outside the BLR when interacting with the infrared photons emitted by the dusty torus. Clearly, a VHE photon of energy $E$ can be absorbed disappearing from the game and producing an electron-positron pair when interacting with an infrared photon of energy $\epsilon$ following the Breit-Wheeler cross-section~\citep{pair1,pair2}
\begin{equation}
\label{cross}
\sigma_{\gamma \gamma}(E,\epsilon,\varphi)=\frac{2\pi\alpha^2}{3m_e^2}W(\beta)~,
\end{equation}
with
\begin{equation}
\label{cross1}
W(\beta)=\left(1-\beta^2\right)\left[2\beta\left(\beta^2-2\right)+\left(3-\beta^4\right)\,{\rm ln}\left(\frac{1+\beta}{1-\beta}\right)\right]~,
\end{equation}
and
\begin{equation}
\label{cross2}
\beta(E,\epsilon,\varphi)\equiv\left[1-\frac{2m_e^2c^4}{E\epsilon(1-{\rm cos}\,\varphi)}\right]^{1/2}~,
\end{equation}
where $\varphi$ is the scattering angle, $\alpha$ is the fine structure constant, $m_e$ is the electron mass and  $c$ is the light velocity. The process is kinematically allowed for
\begin{equation}
\label{cross3}
\epsilon>\epsilon_{\rm thr}(E,\varphi)\equiv\frac{2m_e^2c^4}{E(1-{\rm cos}\,\varphi)}~.
\end{equation}
As in Fig.~\ref{geom} we call $y$ the axis parallel to the torus vertical symmetry axis which is also the jet direction. We take as reference $y=0$, representing the position of the upper plane of the torus: clearly, with this choice the black hole is placed at $y=-h/2$ and we call $y=l=-h/2+r_{\rm em}$ the emission position as in Fig.~\ref{geom}.

In order to obtain the optical depth $\tau(E)$ of VHE photons, we must multiply the spectral number density $n_{\gamma}(\epsilon)$ of background photons with $\sigma_{\gamma \gamma}(E,\epsilon,\varphi)$ and then integrate over the distance, $\epsilon$ and $\varphi$~\citep{tau1,tau2}. As a result, we obtain
\begin{eqnarray}
\label{tau}
&\displaystyle \tau(E)=\int_{y_0}^{\infty}{\rm d}y \int_{({\rm cos}\,\varphi)_{\rm min}(y)}^{({\rm cos}\,\varphi)_{\rm max}(y)}{\rm d}({\rm cos}\,\varphi)\,\frac{1-{\rm cos}\,\varphi}{2} \times \\
&\displaystyle \int_{\epsilon_{\rm thr}(E,\varphi)}^{\infty}{\rm d}\epsilon \, n_{\gamma}(\epsilon)\,\sigma_{\gamma \gamma}(E,\epsilon,\varphi)~, \nonumber
\end{eqnarray}
where
\begin{equation}
\label{nph}
n_{\gamma}(\epsilon)=\frac{\epsilon^2}{\pi^2c^3 \hslash^3}\frac{1}{e^{\frac{\epsilon}{k T}}-1}~,
\end{equation}
is the torus black body spectral number density with $\hslash$ the Planck constant and $k$ the  Boltzmann constant. In addition, for black body emission from the internal vertical surface of the torus we can observe that $y_0=l$, while
\begin{equation}
\label{Lint}
({\rm cos}\,\varphi)_{\rm min}(y)=\frac{y}{(r_{\rm torus, in}^2+y^2)^{1/2}}~,
\end{equation}
and
\begin{equation}
\label{Lint2}
({\rm cos}\,\varphi)_{\rm max}(y)=\frac{h+y}{[r_{\rm torus, in}^2+(h+y)^2]^{1/2}}~.
\end{equation}
Instead, for black body emission from the upper external surface of the torus we have that $y_0=0$, while
\begin{equation}
\label{Lext}
({\rm cos}\,\varphi)_{\rm min}(y)=\frac{y}{(r_{\rm torus, out}^2+y^2)^{1/2}}~,
\end{equation}
and
\begin{equation}
\label{Lext2}
({\rm cos}\,\varphi)_{\rm max}(y)=\frac{y}{(r_{\rm torus, in}^2+y^2)^{1/2}}~.
\end{equation}
This calculation is performed at the redshift $z$ where the FSRQ is located: by calling $E_{\rm obs}$ the energy of the VHE photons as observed at the Earth we must consider cosmological effects so that we have to perform a translation in the new energy reference frame by using the relation $E=E_{\rm obs}(1+z)$.

A VHE photon produced by a FSRQ suffers also absorption because of the interaction with EBL photons in a similar way as discussed above about interaction with photons of the infrared background from the torus. Now, in Eq. (\ref{tau}) we have a redshift dependence of $E$, $\epsilon$ and $n_{\gamma}$ which is presently the EBL spectral number density. However, we directly use the newest data about EBL optical depth~\citep{franc}.

\begin{figure*}       
\begin{center}
\includegraphics[width=.45\textwidth]{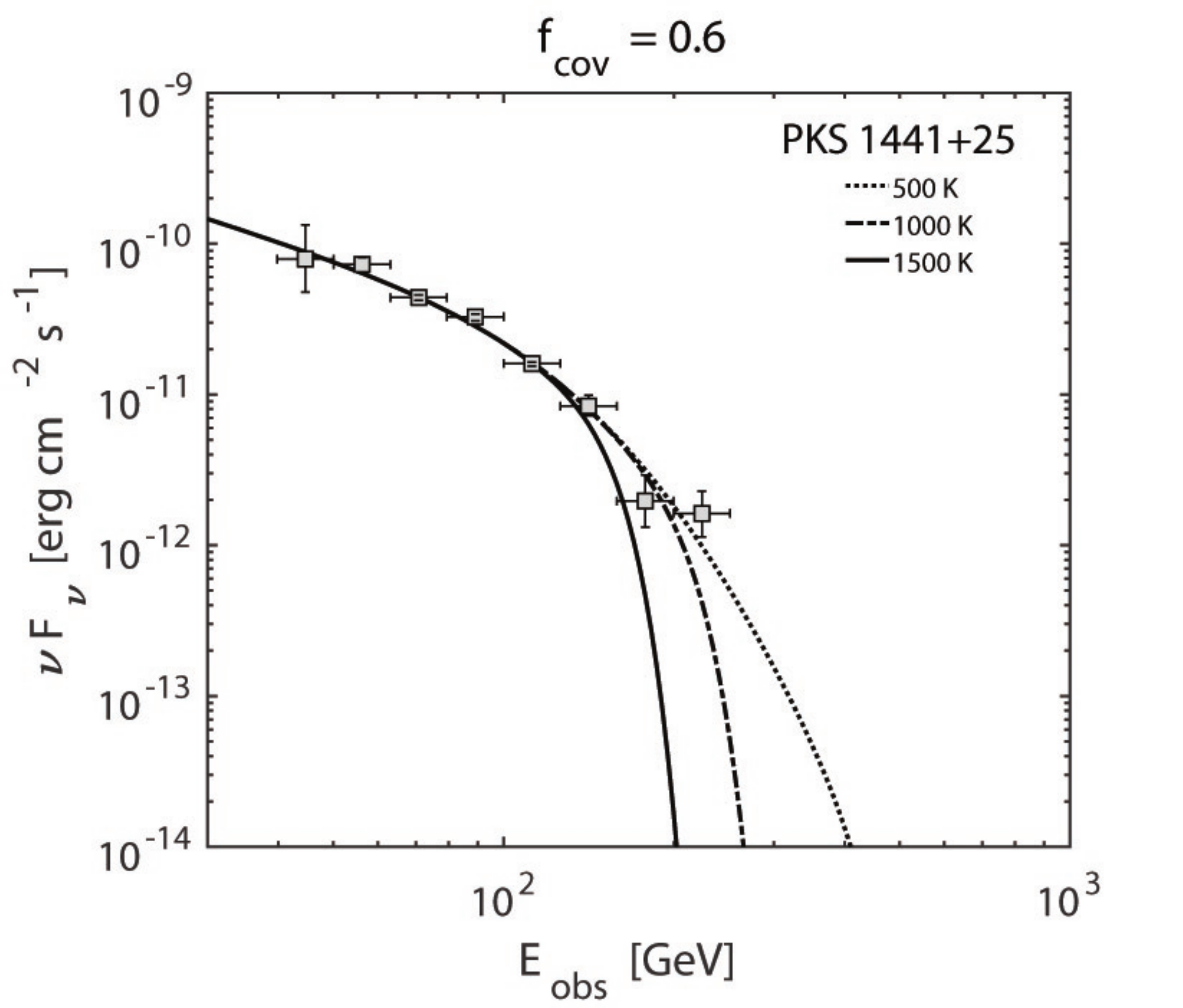}\includegraphics[width=.45\textwidth]{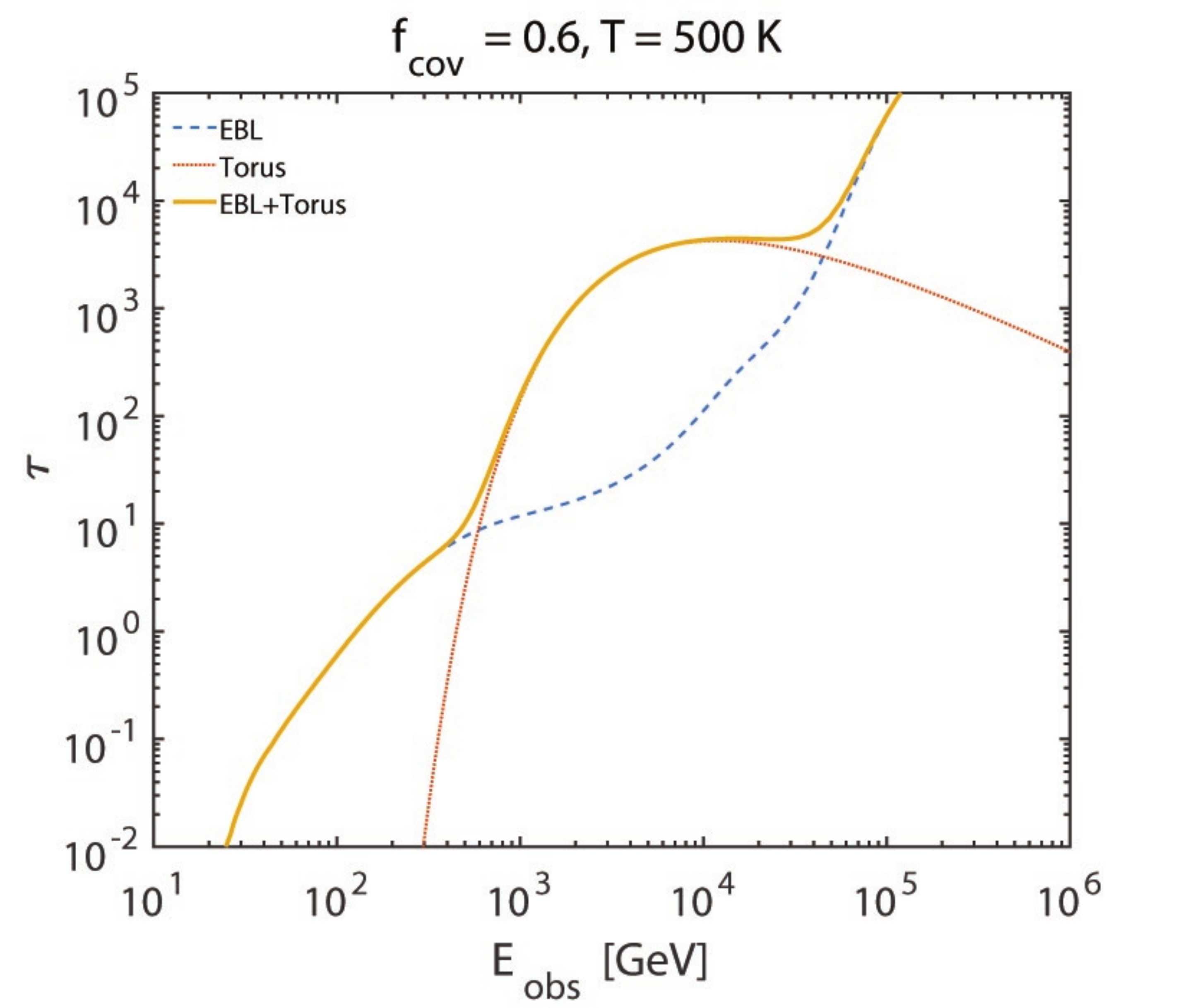}
\includegraphics[width=.45\textwidth]{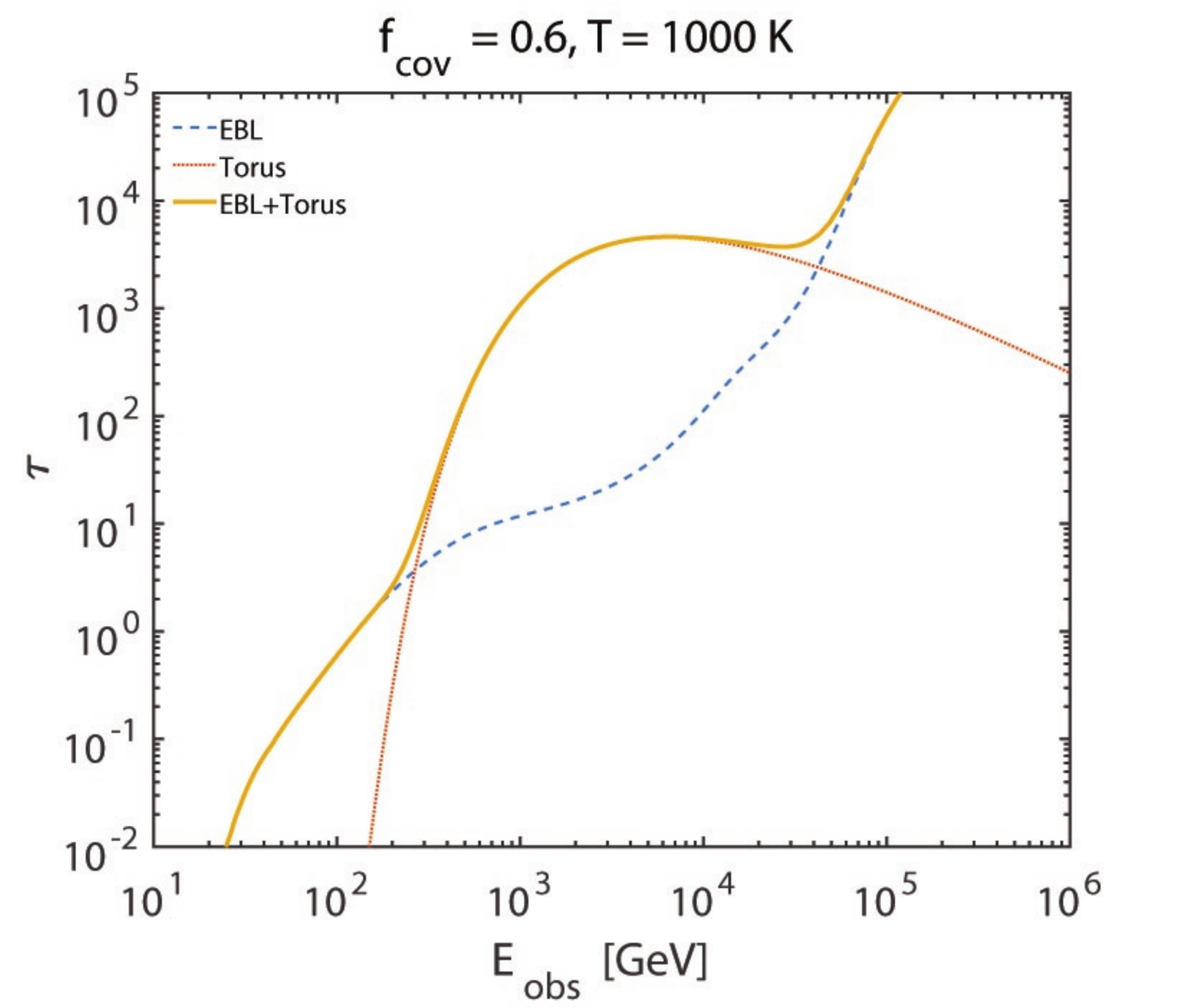}\includegraphics[width=.45\textwidth]{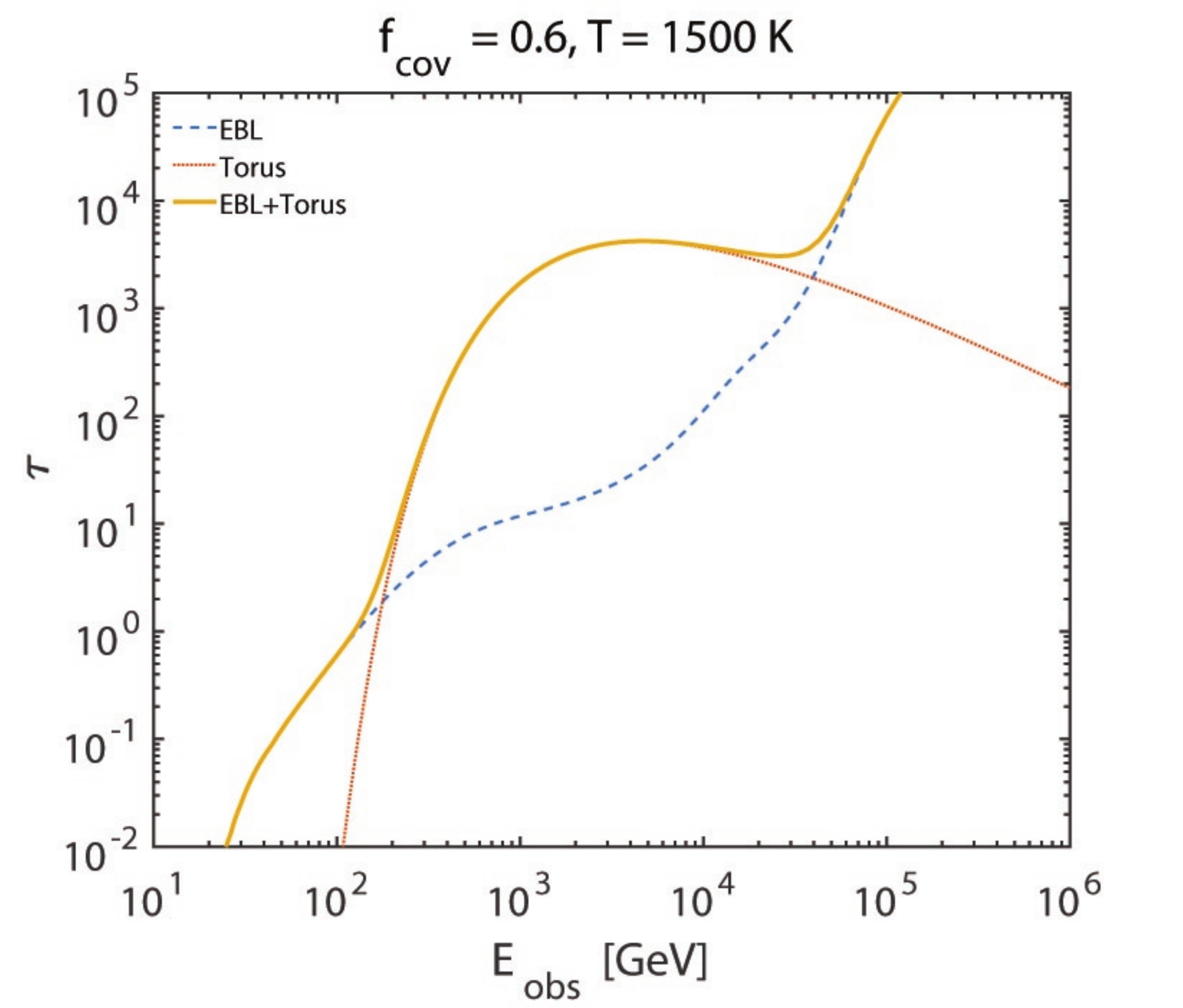}
\end{center}
\caption{\label{temp} 
Behavior of the SED of PKS 1441+25 (upper-left panel) choosing three models for the infrared emission from the torus implying different corresponding torus and total (EBL+torus) optical depths (other three panels). We fix here the value of $f_{\rm cov} =0.6$ and we consider different values for the temperature $T$. The data points are from MAGIC~\citep{PKS1441}.}
\end{figure*}

As an example, we apply our treatment to the FSRQ PKS 1441+25~\citep{PKS1441}. In Fig.~\ref{temp} we explore three different models for the torus: in particular, we fix the covering factor to an intermediate value $f_{\rm cov}=0.6$ and we consider three possible values for the temperature $T=500 \, {\rm K}$, $T=1000 \, {\rm K}$, $T=1500 \, {\rm K}$. From the upper-left panel we infer that it is difficult to distinguish among the different models by using current data (although one could tentatively rule out the model corresponding to $T=1500 \, \rm K$). In particular, the data do not reach energies around $\sim 300 \, {\rm GeV}$ which is the energy where the torus influence starts to become dominant. Therefore, only observational data at higher energy and/or with good enough statistics will likely allow us to understand the torus importance and to discern its properties. The other three panels in Fig.~\ref{temp} show the impact of the torus on the total optical depth that includes also EBL absorption. In particular, we can observe that, as expected, due to the $\gamma\gamma$ threshold (see Eq.~\ref{cross3}), as the temperature increases the peak of maximal absorption shifts at lower and lower energies: this is the reason why in the upper-left panel of Fig.~\ref{temp} the influence of the torus turns out to be at lower and lower energies as the temperature increases.

In Fig.~\ref{fcov} we fix the torus temperature to an intermediate value $T=1000 \, {\rm K}$ and we contemplate three different values for the covering factor $f_{\rm cov} = 0.2$, $f_{\rm cov}= 0.6$ and $ f_{\rm cov}=0.9$. Our findings are similar to the case examined in Fig.~\ref{temp}: again from the upper-left panel we infer that it is difficult to distinguish among the different models with current data for the same reasons expressed above (although the data seem to exclude the extreme value $f_{\rm cov}=0.9$). The other three panels show that as $f_{\rm cov}$ increases, absorption due to the torus becomes bigger and bigger since photons travel longer surrounded by the torus. In particular, the value of the peak becomes huger and huger as $f_{\rm cov}$ rises, while the position of the peak changes less since it is influenced more by the temperature. Even if the behavior of the torus optical depth with respect to the variation of $f_{\rm cov}$ is different compared to the temperature variation, the effects on the SED in the upper-left panel of Figs.~\ref{temp} and~\ref{fcov} are similar and likely indistinguishable.

\begin{figure*}       
\begin{center}
\includegraphics[width=.45\textwidth]{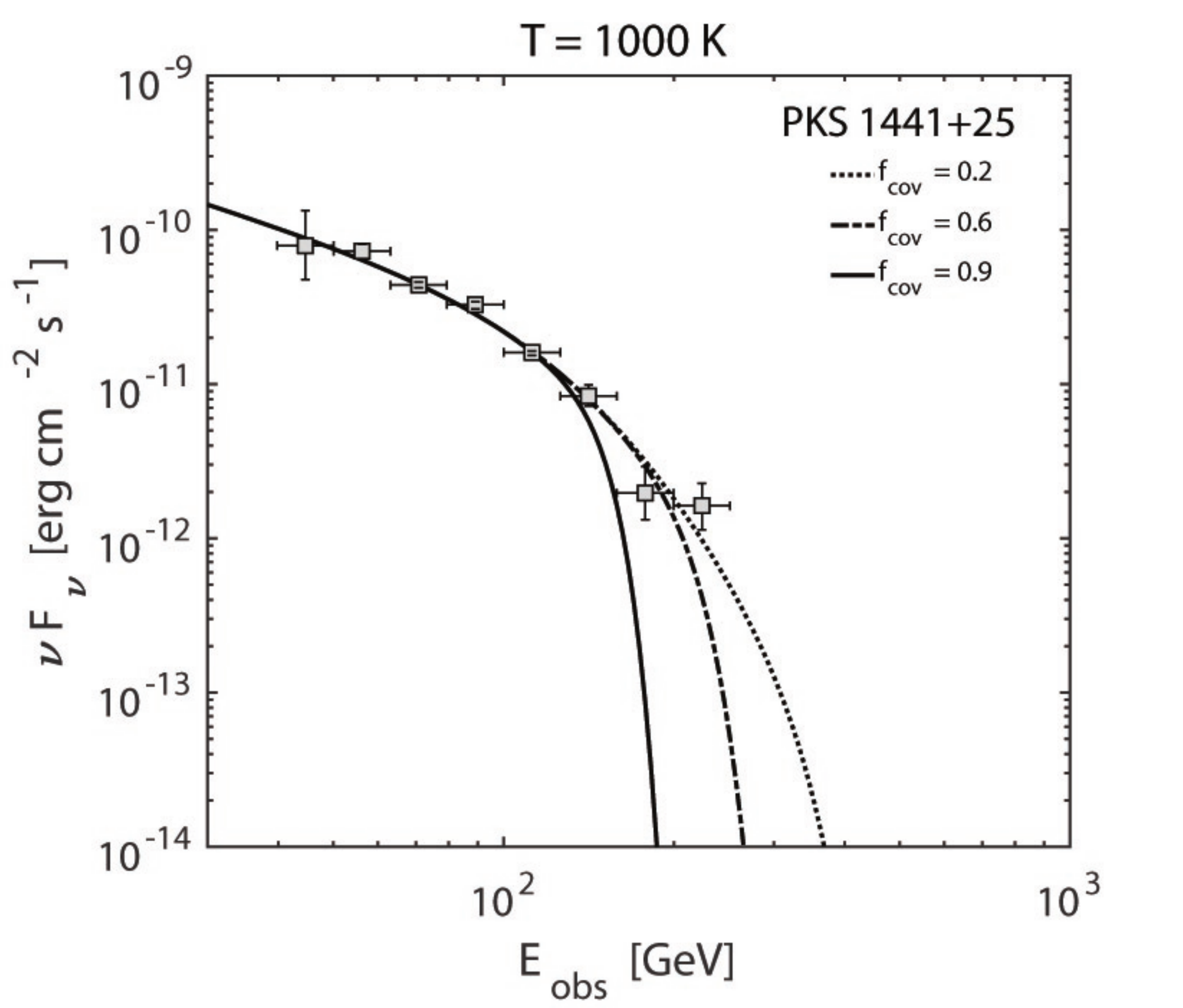}\includegraphics[width=.45\textwidth]{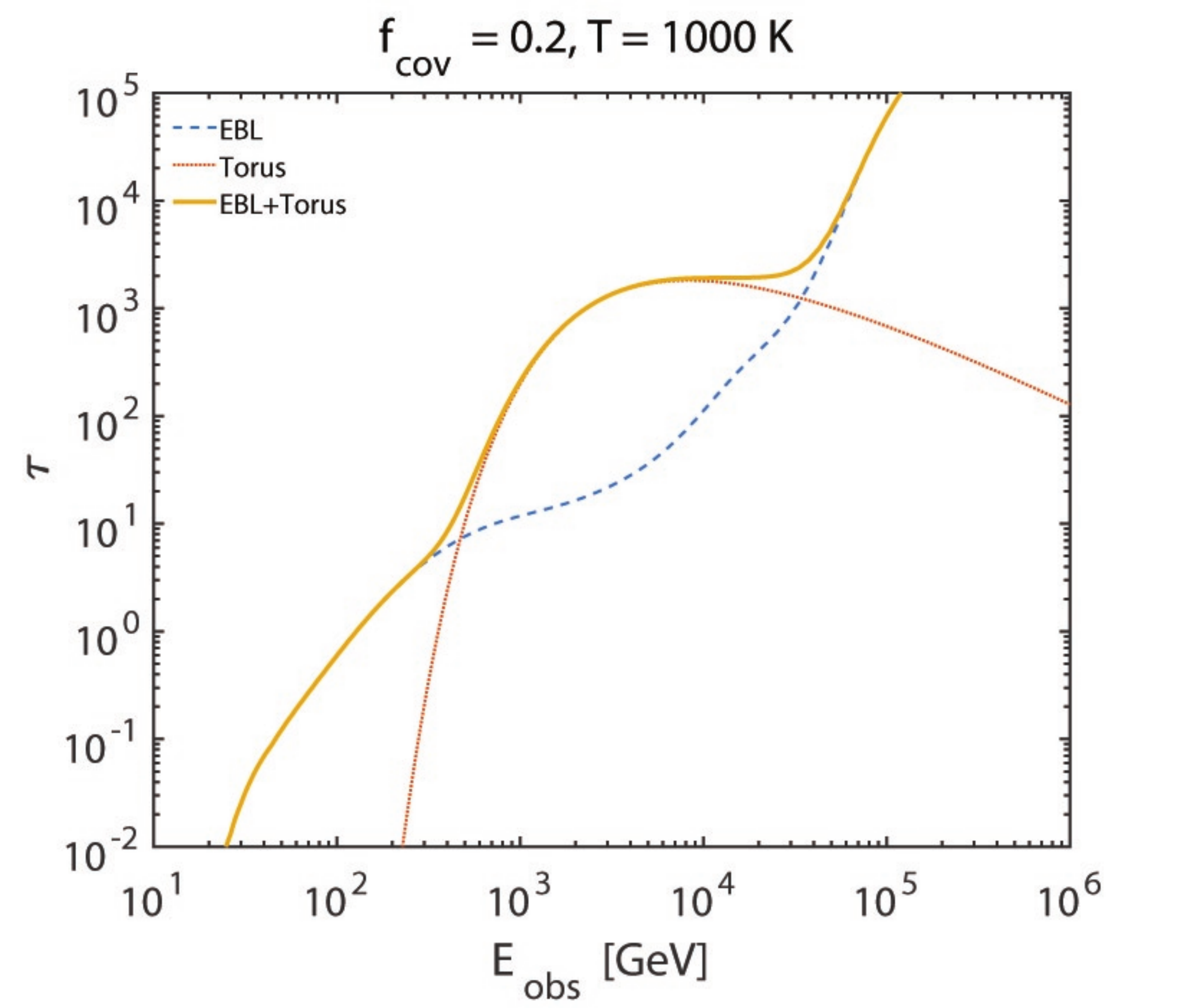}
\includegraphics[width=.45\textwidth]{prova-L2-1000K-f06.pdf}\includegraphics[width=.45\textwidth]{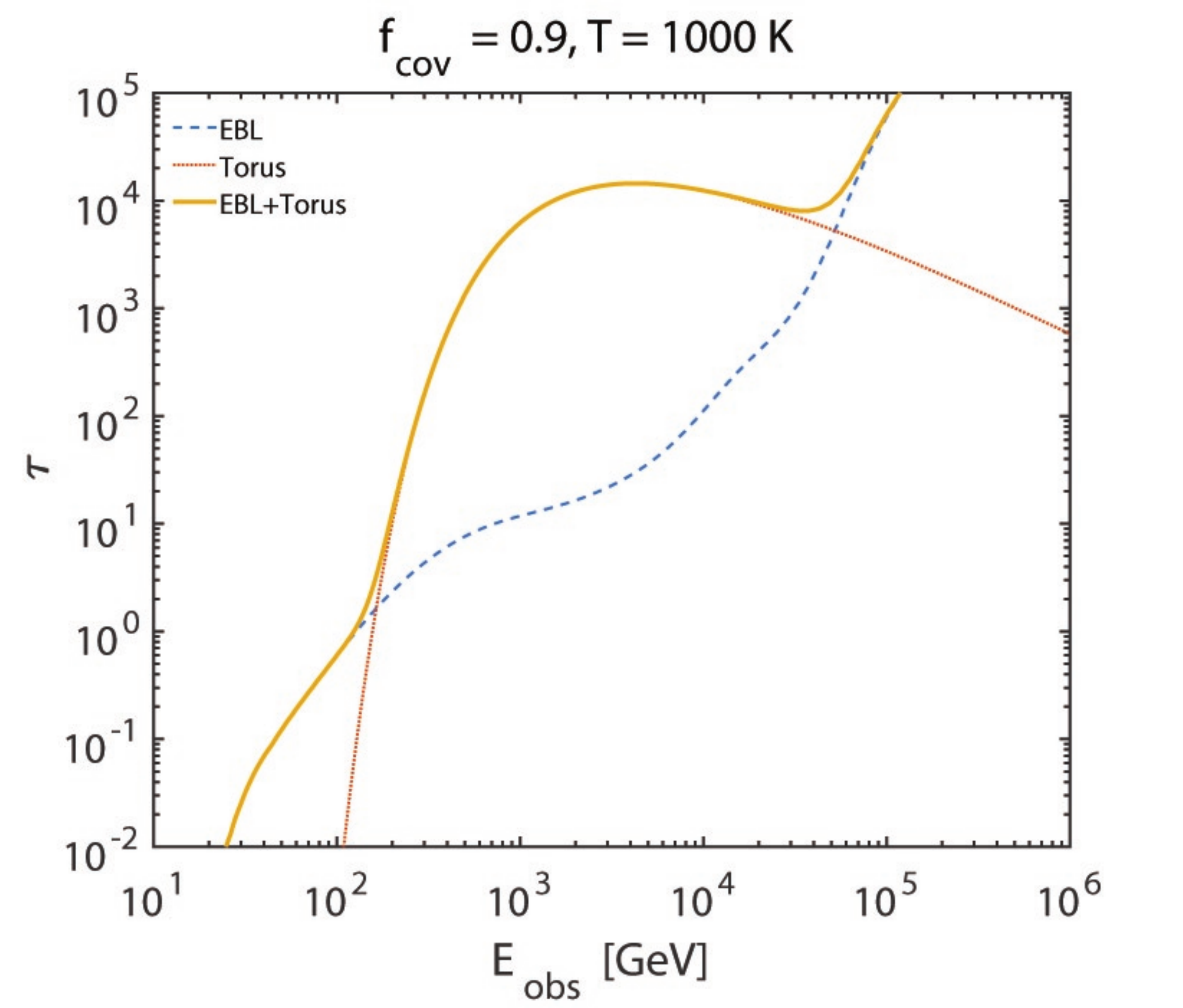}
\end{center}
\caption{\label{fcov} 
Behavior of the SED of PKS 1441+25 (upper-left panel) choosing three models for the infrared emission from the torus implying different corresponding torus and total (EBL+torus) optical depths (other three panels). We fix here the value of $T =1000 \, {\rm K}$ and we consider different values for $f_{\rm cov}$. The data points are from MAGIC~\citep{PKS1441}.}
\end{figure*}

\begin{figure*}       
\begin{center}
\includegraphics[width=.45\textwidth]{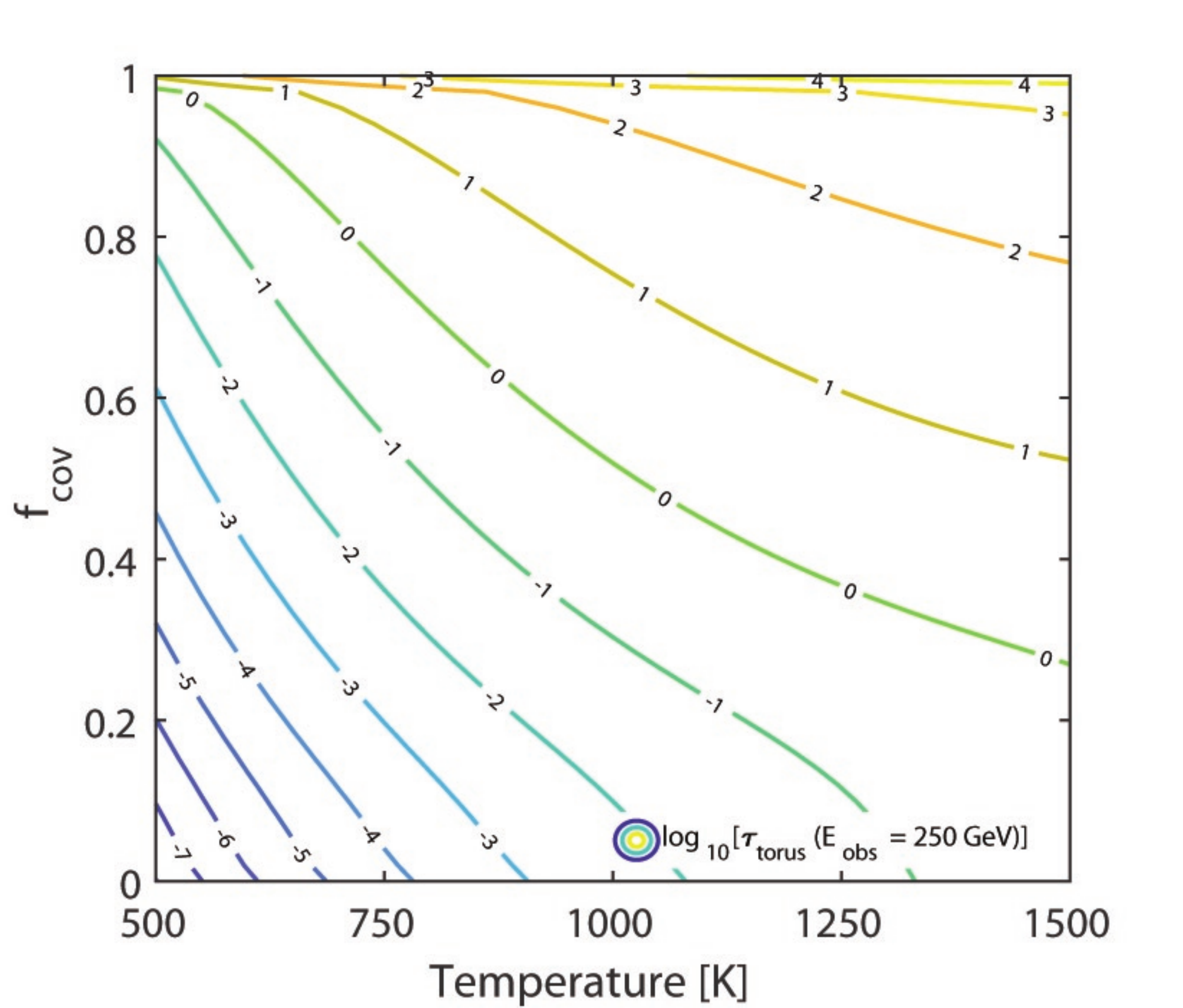}\includegraphics[width=.45\textwidth]{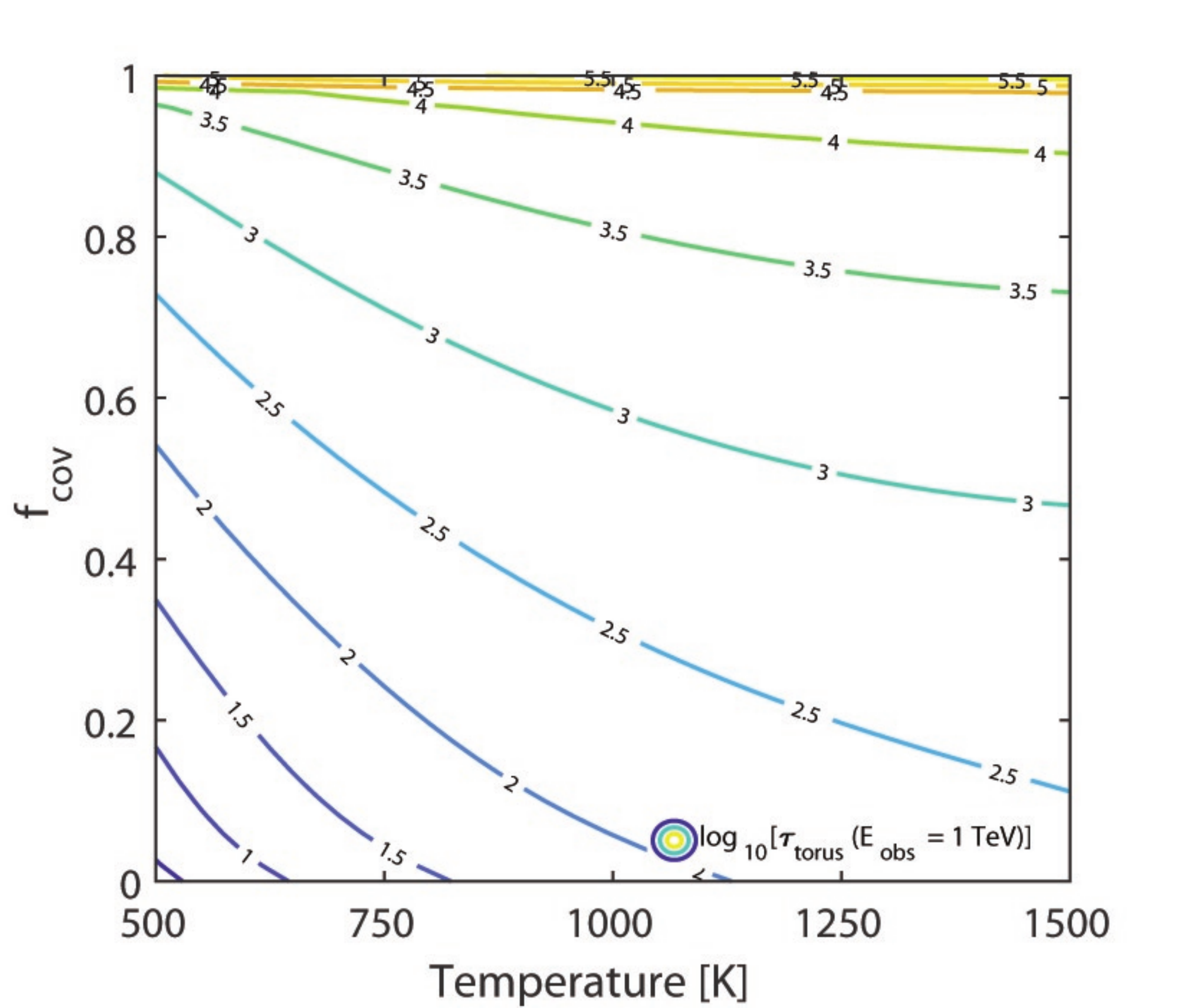}
\includegraphics[width=.45\textwidth]{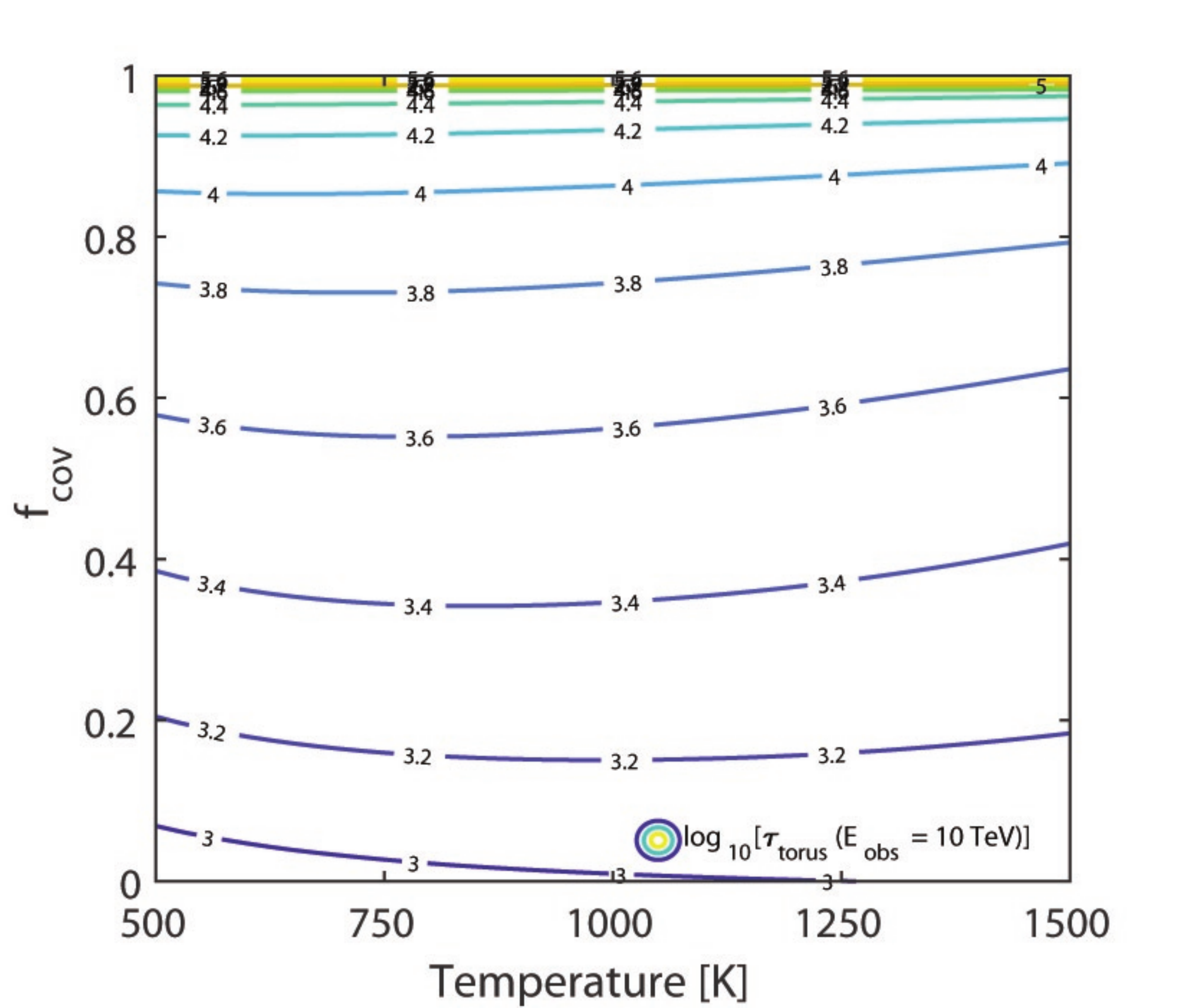}\includegraphics[width=.45\textwidth]{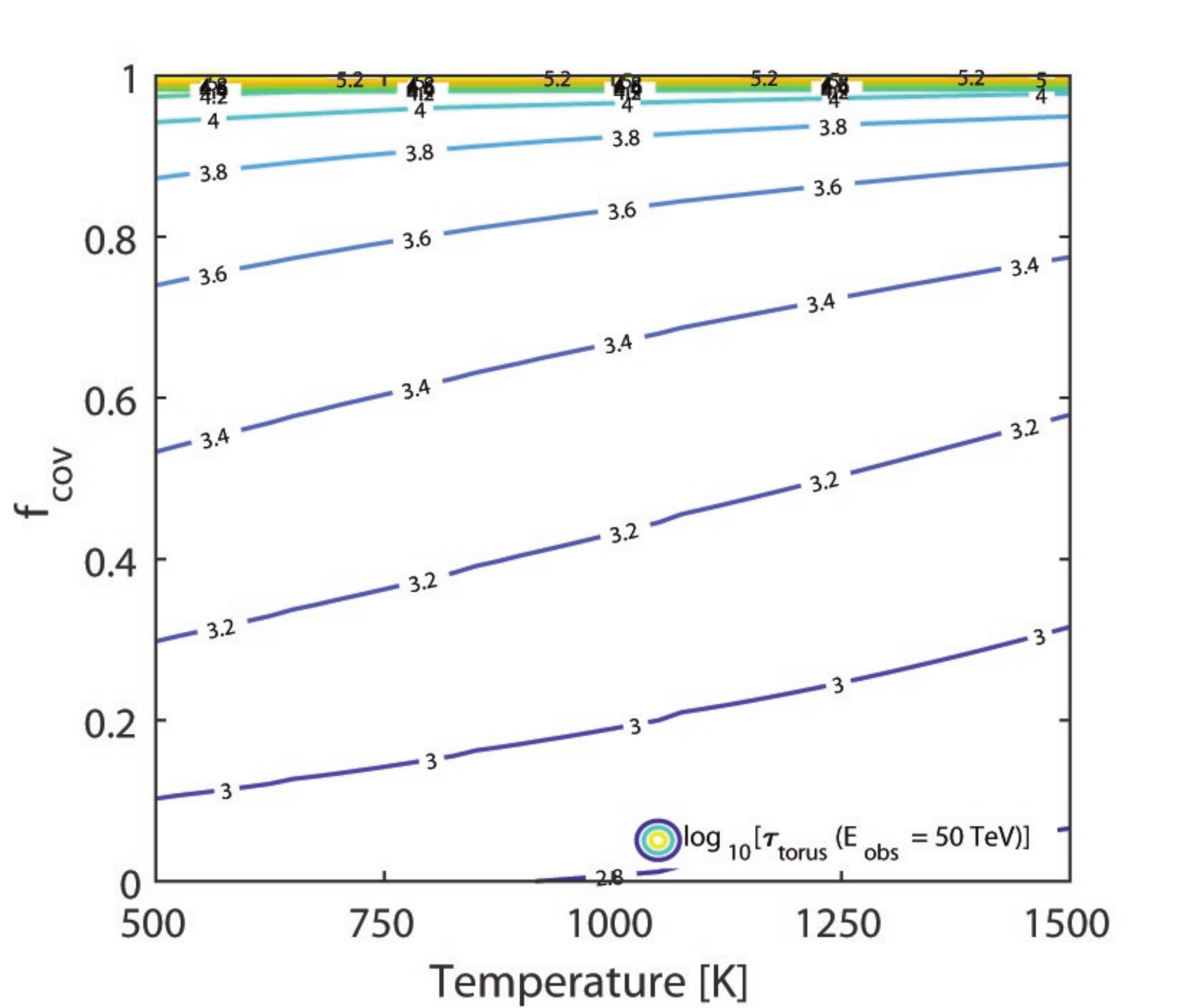}
\end{center}
\caption{\label{degen} 
Torus optical depth contour lines as a function of the temperature and of $f_{\rm cov}$. The behavior of the contour lines changes as the energy increases. The different panels correspond to different choices of the energy of the emitted hard photons.}
\end{figure*}

\begin{figure}       
\begin{center}
\includegraphics[width=.45\textwidth]{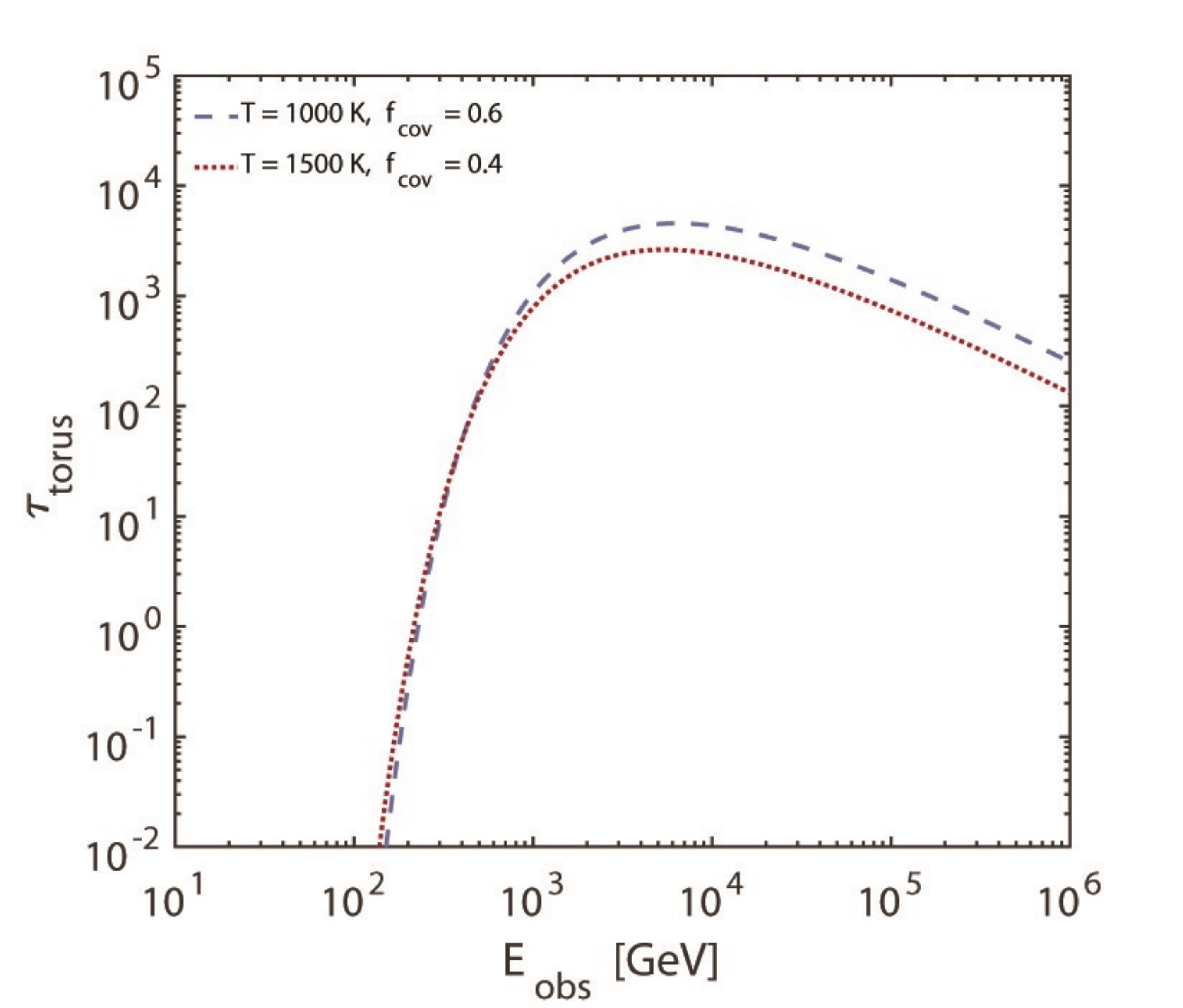}
\end{center}
\caption{\label{confrDeg} 
Torus optical depth $\tau_{\rm torus}$ for (i) $T=1000 \, \rm K$ and $f_{\rm cov}=0.6$ (dashed line) and (ii) $T=1500 \, \rm K$, $f_{\rm cov}=0.4$ (dotted line).}
\end{figure}

As a result we expect a quasi-degeneracy in the torus optical depth: in particular, two different couples of $(T,f_{\rm cov})$ may result in the same value of torus $\tau$. This is indeed what we infer from Fig.~\ref{degen}, where we plot the torus optical depth contour lines as a function of $T$ and $f_{\rm cov}$.
Specifically, we observe a different behavior as the energy increases. At $E_{\rm obs}=250 \, {\rm GeV}$ we observe that in order to have the same value of $\tau$ as $T$ increases $f_{\rm cov}$ must decrease. As the energy increases, the behavior of the contour lines reverses -- this happens around $E_{\rm obs} \sim 1-10 \, {\rm TeV}$. The reason of this change is shown in Figs.~\ref{temp} and~\ref{fcov} if we give a look at the shape of $\tau$: as long as we consider energies smaller than the energy of the torus $\tau$ peak we have that the contour lines behave as in the left- and right-upper panels of Fig.~\ref{degen}, when we explore energies higher than the energy of the torus $\tau$ peak the behavior of the contour lines is inverted. In principle, since the above-defined quasi-degeneracy possesses an energy dependence, an analysis of the torus opacity at different energies might break that degeneracy. However, as plotted in Fig.~\ref{confrDeg} -- where we show two quasi-degenerate cases ($T=1000 \, \rm K$, $f_{\rm cov}=0.6$) and ($T=1500 \, \rm K$, $f_{\rm cov}=0.4$) for $E_{\rm obs}\sim 250 \, \rm GeV$ (their quasi-degeneracy is deductible from the upper-left panel of Fig.~\ref{degen}), the energy where the quasi-degeneracy is broken is above ${\cal O}(1 \, \rm TeV)$ and the opacity is very high with $\tau_{\rm torus}>1000$, so that a detection at those energies appears unlikely. As a result, this quasi-degeneracy transforms {\it de facto} into a degeneracy in the absence of data above ${\cal O}(1 \, \rm TeV)$.

\section{Spectral Energy Distribution (SED)}
The FSRQ photon spectrum can be defined as
\begin{equation}
\label{flux}
{\cal F} (E_{\rm obs}) \equiv \frac{{\rm d}N}{{\rm d}t \, {\rm d}A \, {\rm d}E_{\rm obs}}~,
\end{equation}
where $N$ is the VHE photon number and ${\rm d}A$ is an infinitesimal area. We recall that $E=E_{\rm obs}(1+z)$, where $z$ is the redshift of the FSRQ and $E_{\rm obs}$ is the photon energy as observed at the Earth. We model the intrinsic spectrum of all considered FSRQs with a power law as
\begin{equation}
\label{int}
{\cal F}_{\rm int} (E_{\rm obs}) = {\cal F}_0 \, \left( \frac{E_{\rm obs}}{E_0} \right)^{- k}~,
\end{equation}
with ${\cal F}_0$ a normalization constant accounting for the blazar luminosity, $E_0$ a reference energy and $k$ representing a spectral index. We want to stress that deviations from a pure power law are possible. An additional curvature can be due to the on-set of the Klein-Nishina regime above ${\cal O}(100 \, \rm GeV)$ and/or due to modifications in the electron spectrum. We discuss the effects of this additional curvature and possible disentanglement from that associated to the torus absorption in Sect. 4. The observed FSRQ spectrum is linked to the intrinsic one by
\begin{equation}
\label{obs}
{\cal F}_{\rm obs} (E_{\rm obs}) = e^{-\tau(E_{\rm obs},z)}\, {\cal F}_{\rm int} (E_{\rm obs})~,
\end{equation}
where $\tau$ is the sum of the EBL optical depth $\tau_{\rm EBL}$ and of the torus optical depth $\tau_{\rm torus}$. The SED is related to ${\cal F}_{\rm obs}$ by
\begin{equation}
\label{mr02112018a}
\nu F_{\nu} (E_{\rm obs}) = E_{\rm obs}^2 \, {\cal F}_{\rm obs} (E_{\rm obs})~.
\end{equation}


Now, we can use the theoretical framework developed in Sect. 2 in order to study the spectrum of three FSRQs: PKS 1510-089, PKS 1222+216 and PKS 1441+25. We consider all the FSRQs in flaring state in order to have a more powerful flux and more accurate spectra with better statistics. In addition, in order to avoid the degeneracy given by the couple of parameters $(T,f_{\rm cov})$ we set for all the sources $f_{\rm cov}=0.6$ -- which is considered typical~\citep{fcov}:
\begin{enumerate}
\item {\it PKS 1510-089} -- It is a FSRQ observed at redshift $z=0.361$. We use the observational data points from MAGIC~\citep{PKS1510} which show energies up to $\sim 300 \, {\rm GeV}$. The disk luminosity is estimated to be $L_{\rm disk}\sim 6.7 \times 10^{45} \, {\rm erg \, s^{-1}}$~\citep{PKS1510LUM}, from which, having fixed $f_{\rm cov}=0.6$, we can derive the geometry of the torus as a function of the temperature $T$ only. In particular, $r_{\rm torus,in}$ directly reads from Eq. (\ref{t1}), as mentioned above $r_{\rm torus,out}=2 \, r_{\rm torus,in}$, $h$ is obtained from Eq. (\ref{t2}) and $r_{\rm BLR}$ from Eq. (\ref{t4}). In order to obtain the SED we take $E_0 = 100 \, {\rm GeV}$ and $k = 3.0$ in Eq. (\ref{int}).



\item {\it PKS 1222+216} -- This FSRQ is located at redshift $z=0.432$. The observational data points from MAGIC~\citep{PKS1222} reach energies up to $\sim 350 \, {\rm GeV}$. We consider a disk luminosity $L_{\rm disk}\sim 1.5 \times 10^{46} \, {\rm erg \, s^{-1}}$~\citep{PKS1222LUM}. The geometry of the torus can be inferred by using the same procedure as in the case of PKS 1510-089. The SED in Eq. (\ref{int}) is calculated by taking $E_0 = 100 \, {\rm GeV}$ and $k = 2.4$.



\item {\it PKS 1441+25} -- It is a very distant FSRQ observed at redshift $z=0.940$. This FSRQ has been observed by MAGIC~\citep{PKS1441} with energies up to $\sim 250 \, {\rm GeV}$. We assess a disk luminosity of $L_{\rm disk}\sim 2 \times 10^{45} \, {\rm erg \, s^{-1}}$~\citep{PKS1441LUM}. We can define the torus geometry by using the same strategy performed for PKS 1510-089. In Eq. (\ref{int}) we take $E_0 = 100 \, {\rm GeV}$ and $k = 3.1$ in order to calculate the SED.



\end{enumerate}

\begin{figure*}       
\begin{center}
\includegraphics[width=.33\textwidth]{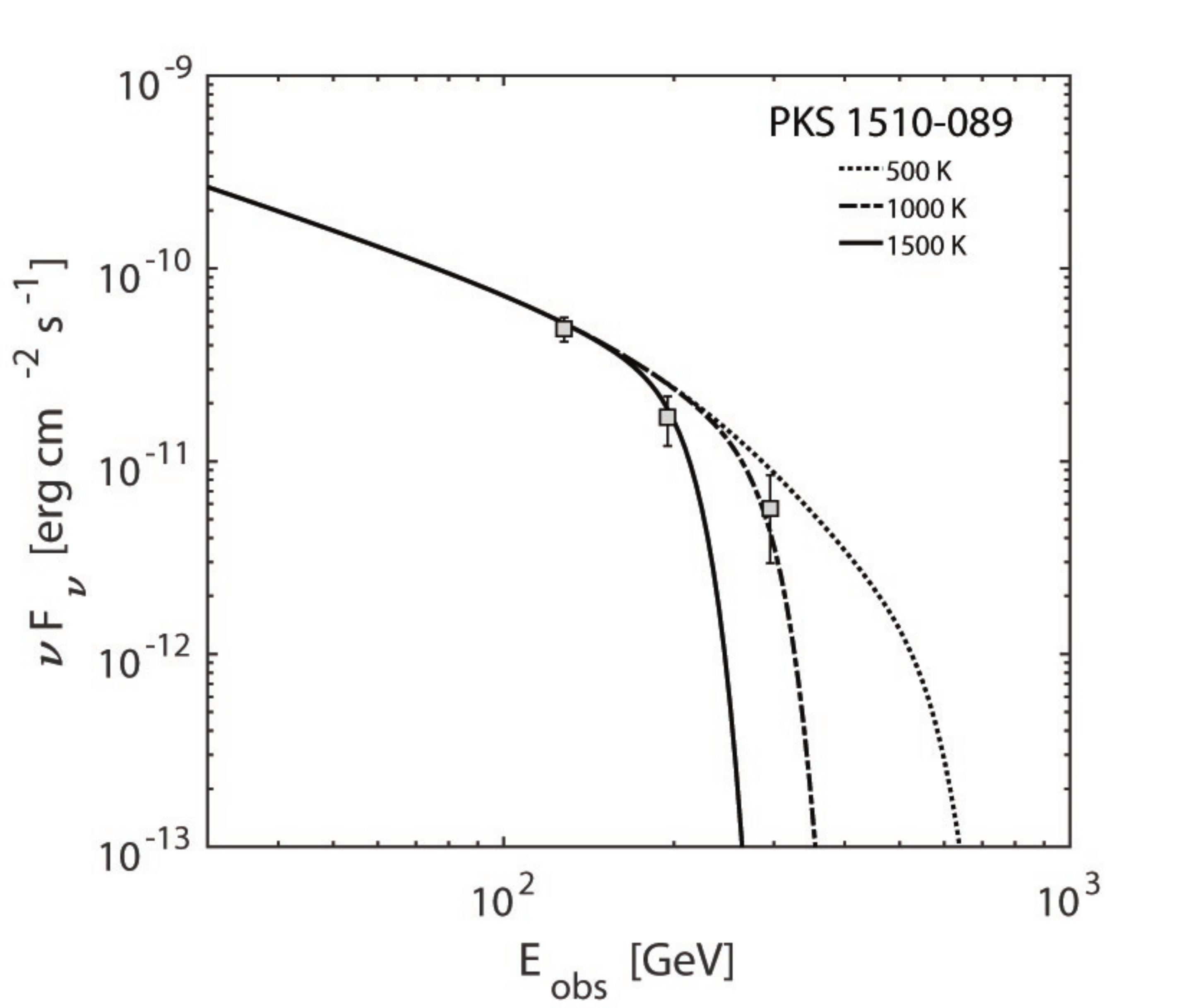}\includegraphics[width=.33\textwidth]{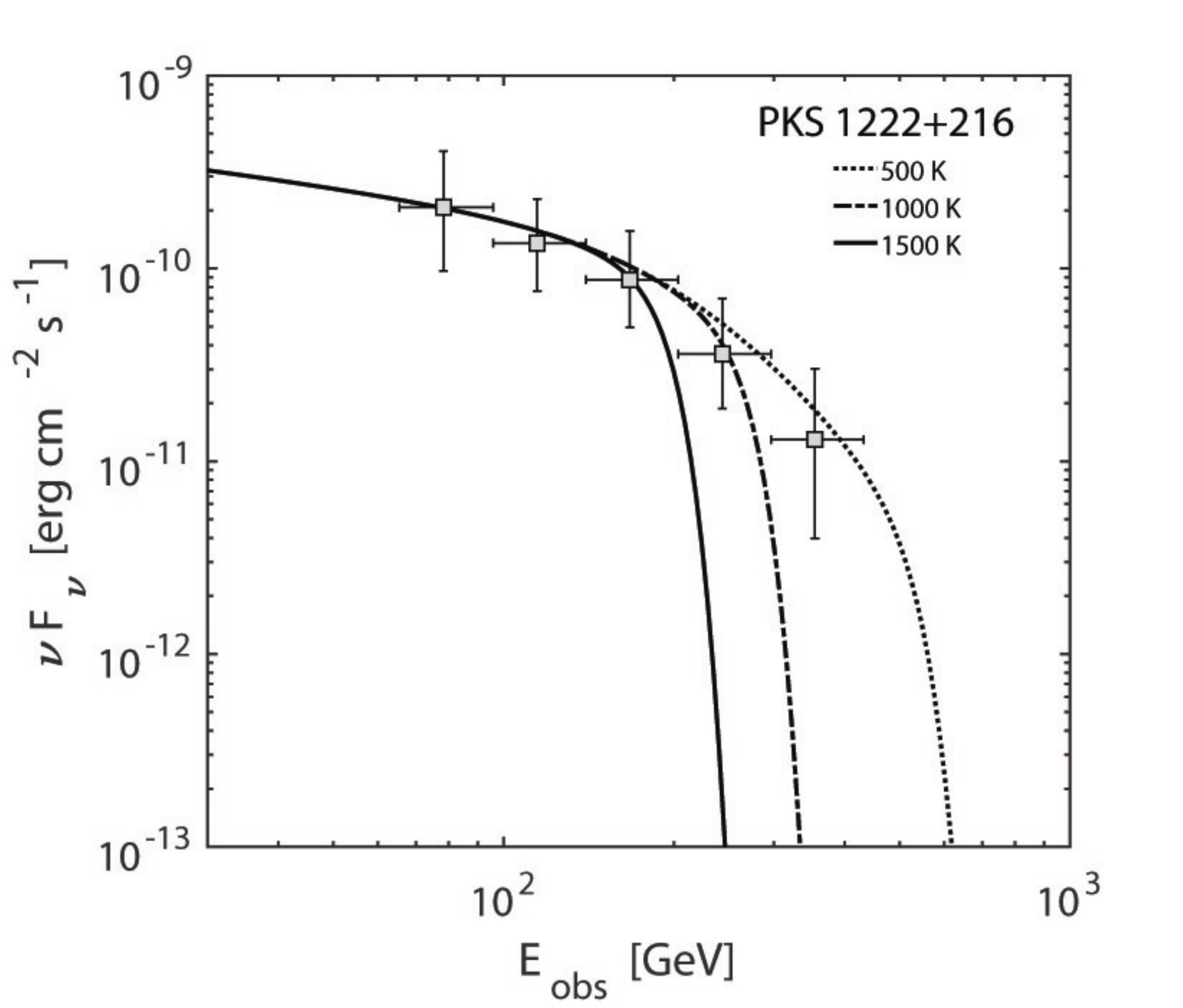}\includegraphics[width=.33\textwidth]{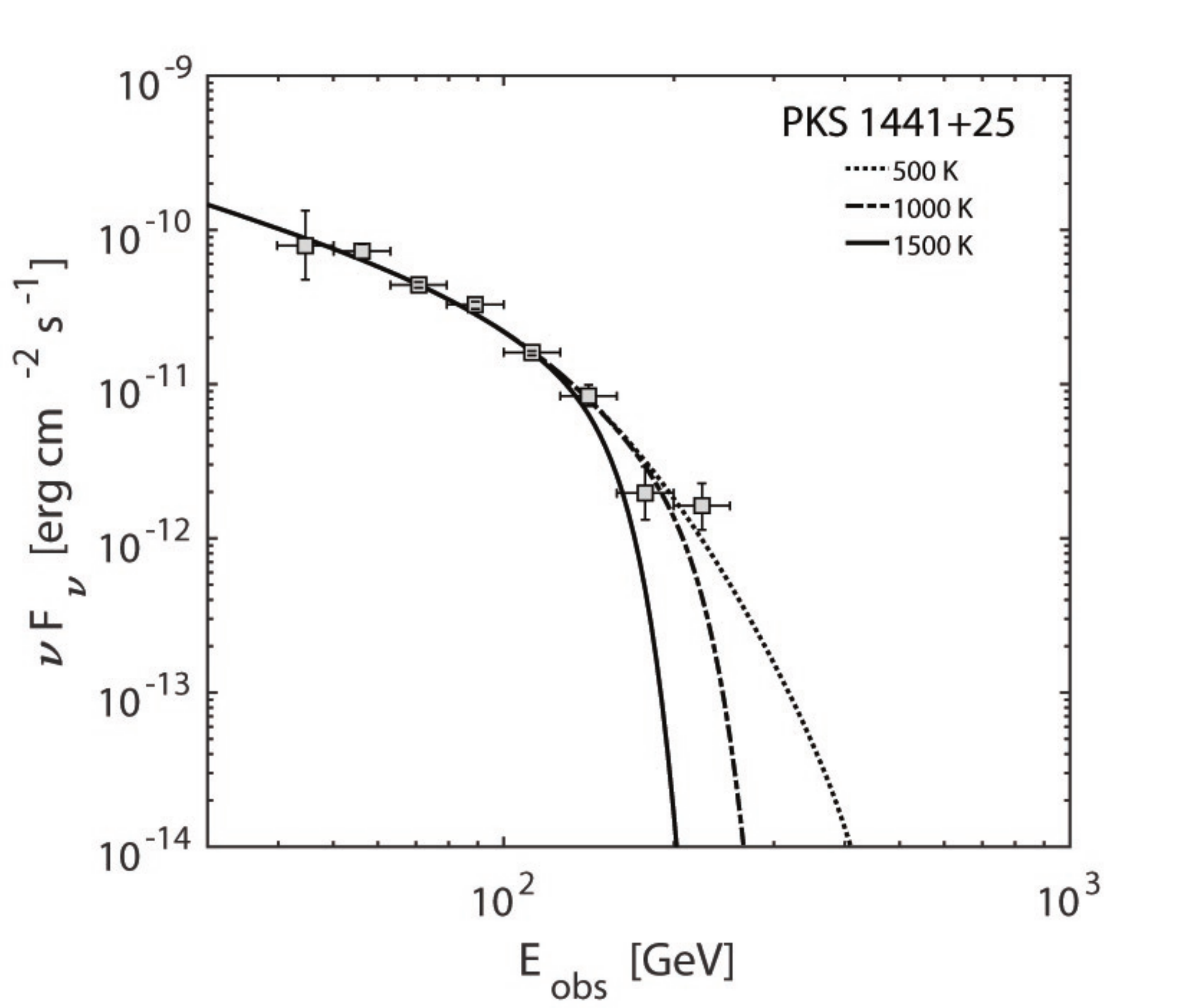}
\end{center}
\caption{\label{SEDall} 
SED of PKS 1510-089 (left panel), PKS 1222+216 (central panel) and  PKS 1441+25 (right panel). We take a fixed $f_{\rm cov}=0.6$. The dotted black line corresponds to a torus model with $T=500 \, {\rm K}$, the dotted-dashed black line is referred to $T=1000 \, {\rm K}$ and the solid black line to $T=1500 \, {\rm K}$. The data points are all from MAGIC: \citet{PKS1510} for PKS 1510-089, \citet{PKS1222} for PKS 1222+216 and \citet{PKS1441} for PKS 1441+25.
}
\end{figure*}

In Fig.~\ref{SEDall} we plot the SED of all considered FSRQs: we plot three models for the torus with different temperatures, $T=500 \, {\rm K}$, $T=1000 \, {\rm K}$ and $T=1500 \, {\rm K}$.
In all the panels of Fig.~\ref{SEDall} we draw the observed SED calculated with the torus optical depth derived by Eq. (\ref{tau}) combined with the EBL optical depth in order to obtain Eq. (\ref{mr02112018a}).  

Fig.~\ref{SEDall} shows that it is currently hardly possible with present day observational data to distinguish among the different torus models since present data do not arrive at energies where the torus influence is prominent. However, Fig.~\ref{SEDall} indicates that a temperature $T=1500 \, \rm K$ is rather disfavored. Nevertheless, current data are not precise enough to assess a strong conclusion, so that only forthcoming higher energy CTA data will likely clarify the torus importance and will determine its properties.

\section{Simulated spectra}
In the previous sections we have taken the torus temperature $T$ in the range $500 \, {\rm  K}-1500 \, {\rm K}$ and convering factor $f_{\rm cov}$ in the range $0.2-0.9$ for illustrative purposes in order to show more evidently how $T$ and $f_{\rm cov}$ affect the observed spectrum. In the following, we modify the torus temperature range raising its lower limit to $T=750 \, \rm K$ since lower values are considered as disfavored (see~\citealt{fcov}) and increasing its upper limit to $T=1750 \, \rm K$ (a physically motivated possible value) in order to test CTA sensitivity in detecting torus temperature. In addition, we fix $f_{\rm cov}$ to its typical value $f_{\rm cov}=0.6$ (see~\citealt{fcov}). We use a simple power law as intrinsic spectrum: possible deviations from a pure power law are due to the on-set of the Klein-Nishina regime above ${\cal O}(100 \, {\rm GeV})$ and/or due to modifications in the electron spectrum. First, we have calculated the spectral models by following the procedure developed in the previous sections. The absorption due to the torus is obtained by means of the model developed in Sect. 2. Once we have added the contribution of the EBL absorption, we can derive the observed FSRQ spectra and simulate the observational data bins observable by the CTA. We closely follow the same prescriptions established in~\citet{HBsimu}, so that we use the analysis package for IACT data \texttt{CTOOLS}\footnote{http://cta.irap.omp.eu/ctools/}~\citep{knodl}, and the public CTA instrument response function\footnote{https://www.cta-observatory.org/science/cta-performance/} (IRF). In particular, we have used version 1.6.1 of ctools suite and proper IRFs according to the site (North or South), zenith angle and exposure time (50h-IRFs for the cases with $50 \, \rm h$ of exposure and 5h-IRFs for the ones with $10 \, \rm h$, which is the closest provided for simulations). Since the computational burden of simulating the spectra is remarkable, we use the procedure developed in \citet{landoni1, landoni2} and already successfully tested and validated on various scientific cases (e.g. \citealt{HBsimu, romano}).
In our dedicated simulations we consider the exact source position in the sky, the correct corresponding CTA IRF and we assume 10 and 50 h of exposure. We obtain the observed spectrum energy bins that are constructed according to the CTA energy resolution~\citep{CTAsens} and by considering the strong photon flux expected by the three analyzed FSRQs. 

\begin{figure}       
\begin{center}
\includegraphics[width=.45\textwidth]{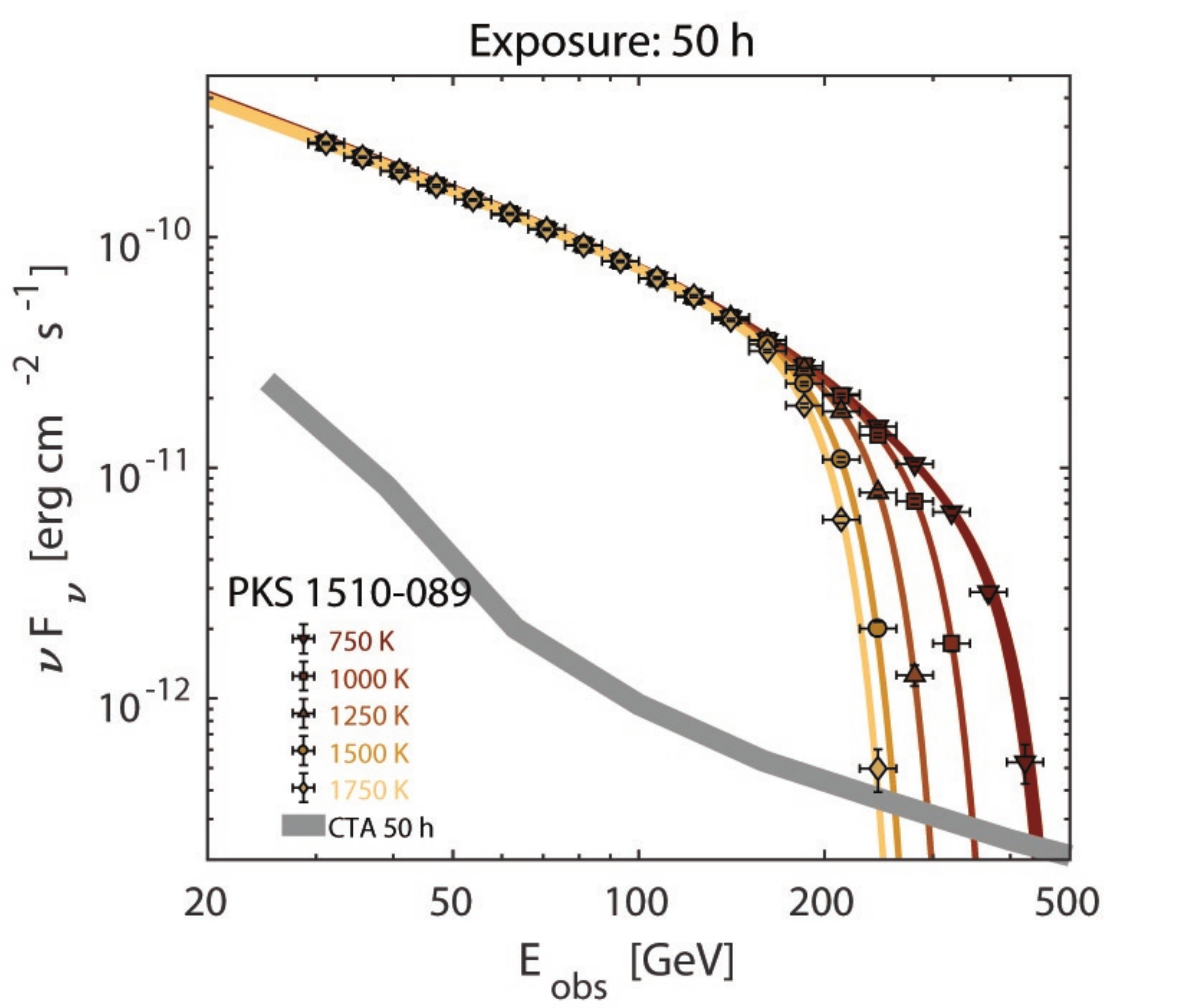}
\includegraphics[width=.45\textwidth]{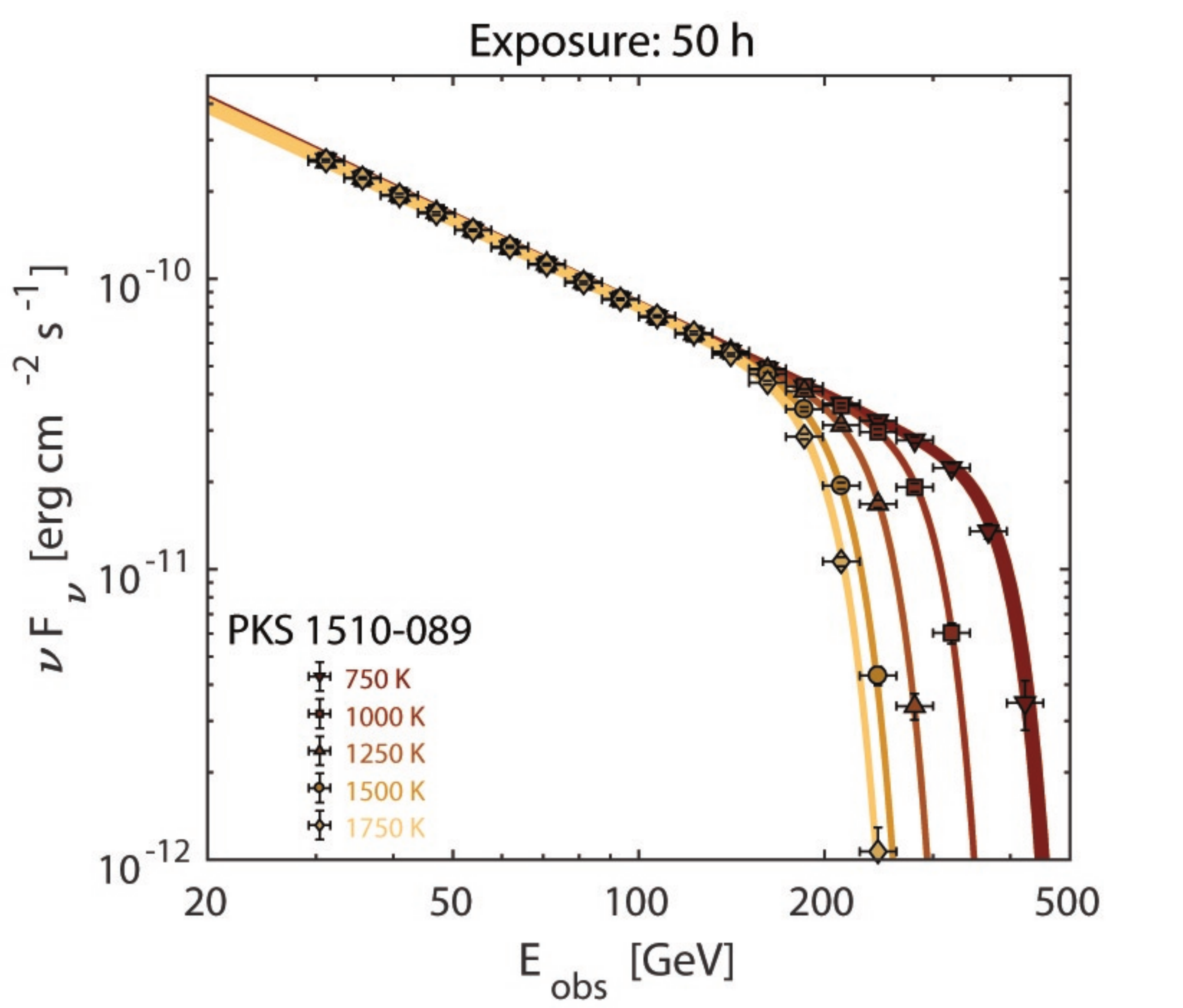}
\end{center}
\caption{\label{bins1510obs} 
Spectrum energy bins of PKS 1510-089 in case of 50 h of observation with different choices of the torus temperature. In the upper panel we report the observed spectrum, while the same spectrum once EBL-deabsorbed is put in the lower panel. The solid gray line represents the CTA sensitivity for 50 h of observation~\citep{CTAsens}.}
\end{figure}

We now perform a statistical analysis on the simulated spectrum energy bins of all three considered FSRQs once their spectrum is EBL-deabsorbed. In Fig.~\ref{bins1510obs} we report as an example on the top panel the observed spectrum energy bins for PKS 1510-089 and in the lower panel the same spectrum energy bins but EBL-deabsorbed.
In the lower panel of Fig.~\ref{bins1510obs} we can recognize the PL behaviour at low energies and the influence of the torus starting from $\sim 200 \, {\rm GeV}$. Then, we try to fit the EBL-deabsorbed spectrum by means of two fitting functions: (i) a power law (PL) of equation
\begin{equation}
\label{PLfit}
F(E_{\rm obs})=K\left(\frac{E_{\rm obs}}{E_0}\right)^{\gamma}~,
\end{equation}
where $K$ is the normalization constant, $E_0$ is a reference energy and $\gamma$ is the spectral index, and (ii) a super exponential cut-off power law (SPL) of equation
\begin{equation}
\label{SPLfit}
F(E_{\rm obs})=K\left(\frac{E_{\rm obs}}{E_0}\right)^{\gamma}e^{-\left(\frac{E_{\rm obs}}{E_c}\right)^{\alpha}}~,
\end{equation}
where $E_{c}$ is the cut-off energy and $\alpha$ is the super exponential index. When the PL of Eq.~(\ref{PLfit}) is considered, we do not have any information about the torus properties, namely its temperature, while when we adopt the SPL model of Eq.~(\ref{SPLfit}) we observe that $E_c$ is strictly linked to the torus temperature. Eq.~(\ref{SPLfit}) represents a phenomenological approximation to the torus model developed in the previous sections: still, the SPL model is very effective in reproducing torus effects in observed data. As a result, by using future observational data we expect that it will be possible to detect $E_c$ with a rather good accuracy and, since there exists a strict relation between $E_c$ and $T$, to distinguish among different torus temperature ranges provided that other torus parameters are constrained enough. Of course, real data can also be directly fitted by the physical models calculated in the previous sections and, as a consequence, the temperature can be directly deduced by the models themselves.

\begin{figure}       
\begin{center}
\includegraphics[width=.45\textwidth]{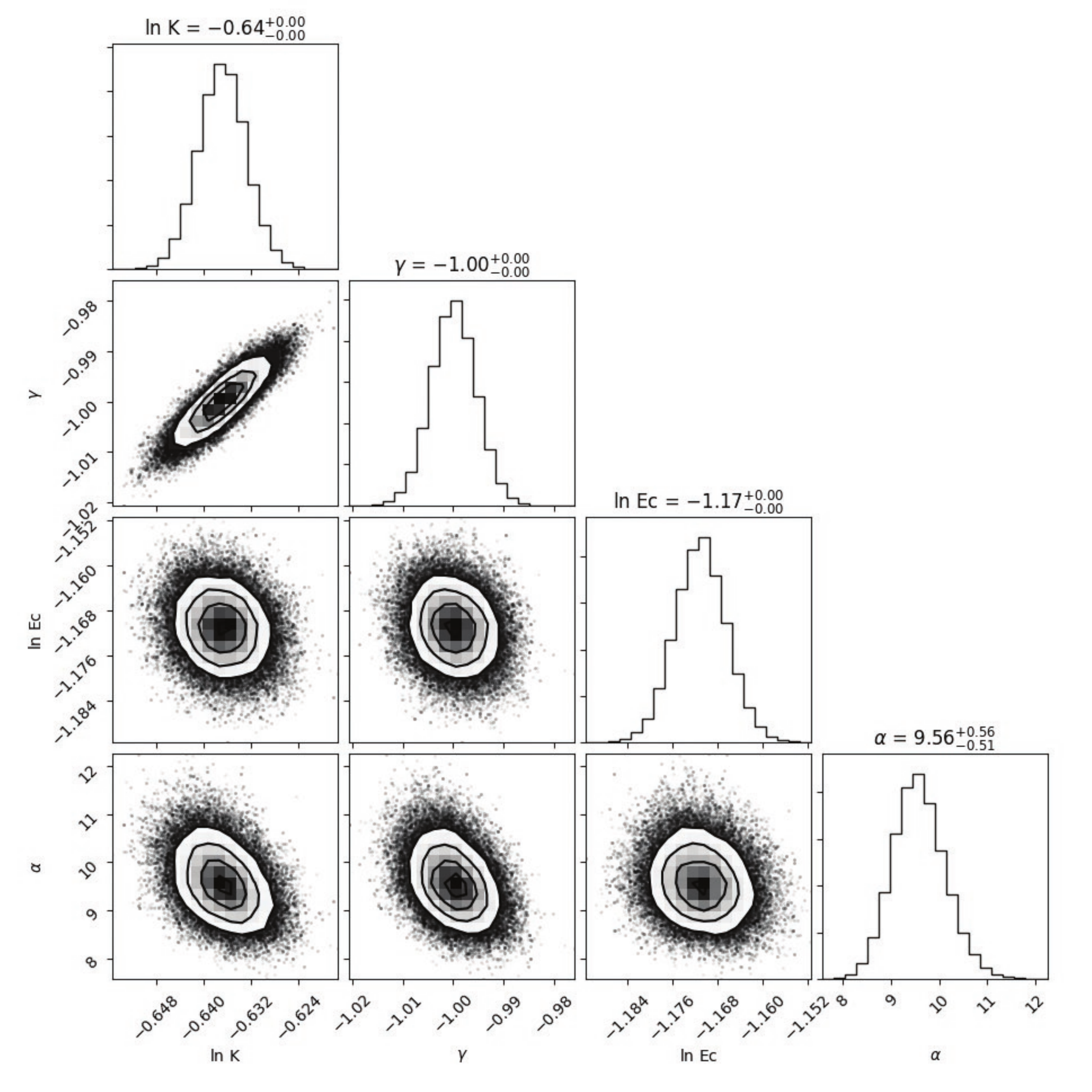}
\end{center}
\caption{\label{cornerPlot} 
Corner plot for the super exponential cut-off power law (SPL) model for PKS 1510-089 in case of 50 h of observation and a torus temperature $T=1000 \, \rm K$.}
\end{figure}

In order to infer which model between PL and SPL better describes the data we perform a Bayesian analysis~\citep[e.g. ][]{KassRaftery1995,Ivezicetal2014} for all the FSRQs at all the considered torus temperatures: 750 K, 1000 K, 1250 K, 1500 K, 1750 K and for 10 h and 50 h of observation. Just as an example, we report in Fig.~\ref{cornerPlot} the corner plot for the SPL model for PKS 1510-089 with torus temperature $T=1000 \, \rm K$ and 50 h of observation.
The procedure implies to fit the available spectra with the two models maximizing the likelihood function by a non-linear optimization algorithm \citep[e.g. the Nelder-Mead algorithm,][]{GaoHan2012} and integrating the posterior probability density of the parameters of our models by a Markov Chain Monte Carlo \citep[MCMC,][]{HoggForeman2018} based on the ``parallel-tempering ensemble'' algorithms \citep{Foreman-Mackeyetal2013}. 
We started the chains from small Gaussian balls centered on the best fit values. The first third of each chain (the ``burn-in phase'') was discarded and we checked that a stationary distribution was reached \citep{Sharma2017}. 
Model comparison can be carried out computing the posterior probabilities by the so-called ``thermodynamic integration'' \citep{GoggansChi2004}. If the posterior probabilities of the models are, respectively, $p_{1}$ and $p_{2}$, Bayes factors are simply given by $p_2/p_1$, and they can be easily converted to probabilities conditioned on the data in favor of the second model with respect the first one, as $p = p_2/(p_1+p_2)$.

\begin{figure*}       
\begin{center}
\includegraphics[width=.33\textwidth]{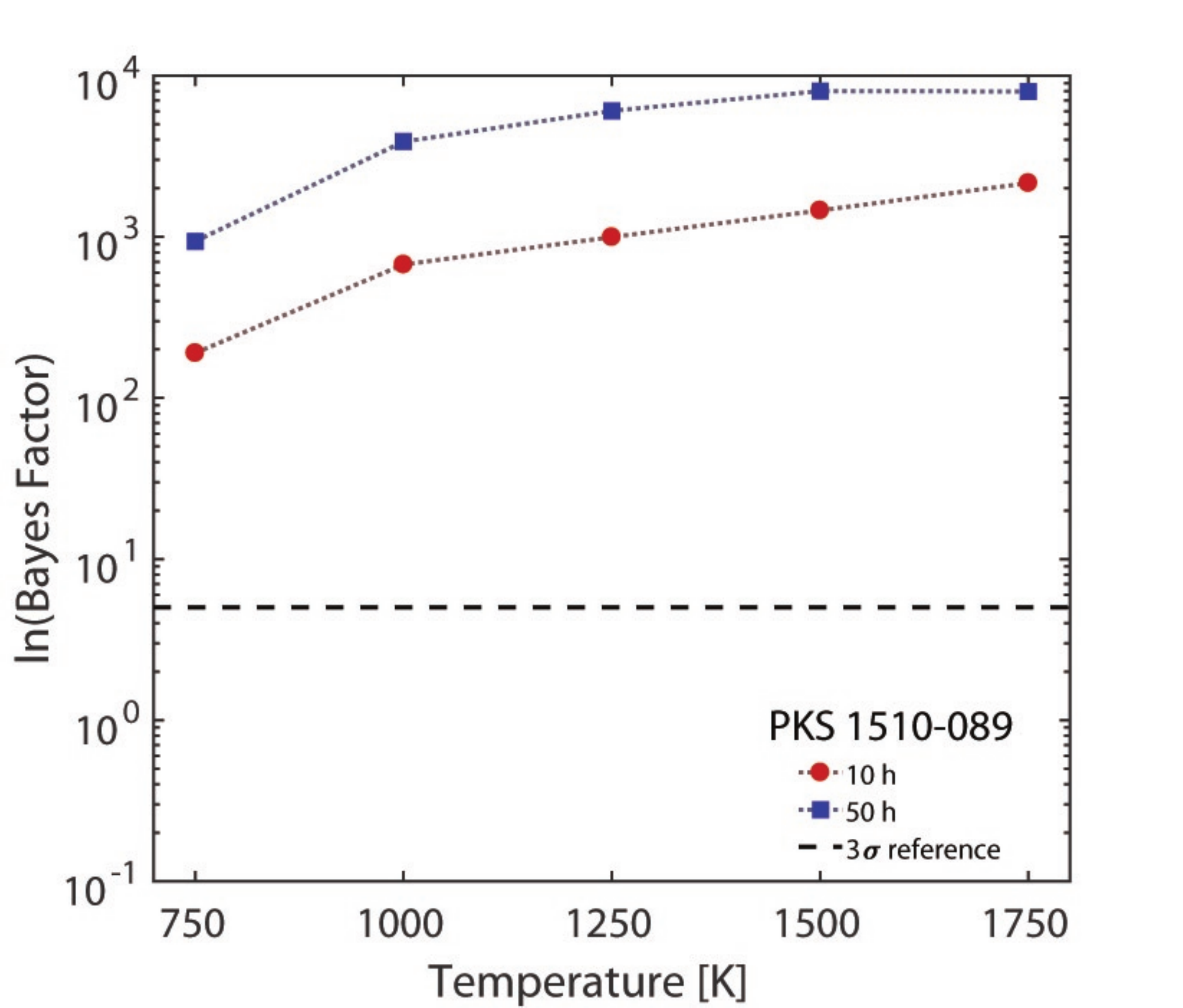}\includegraphics[width=.33\textwidth]{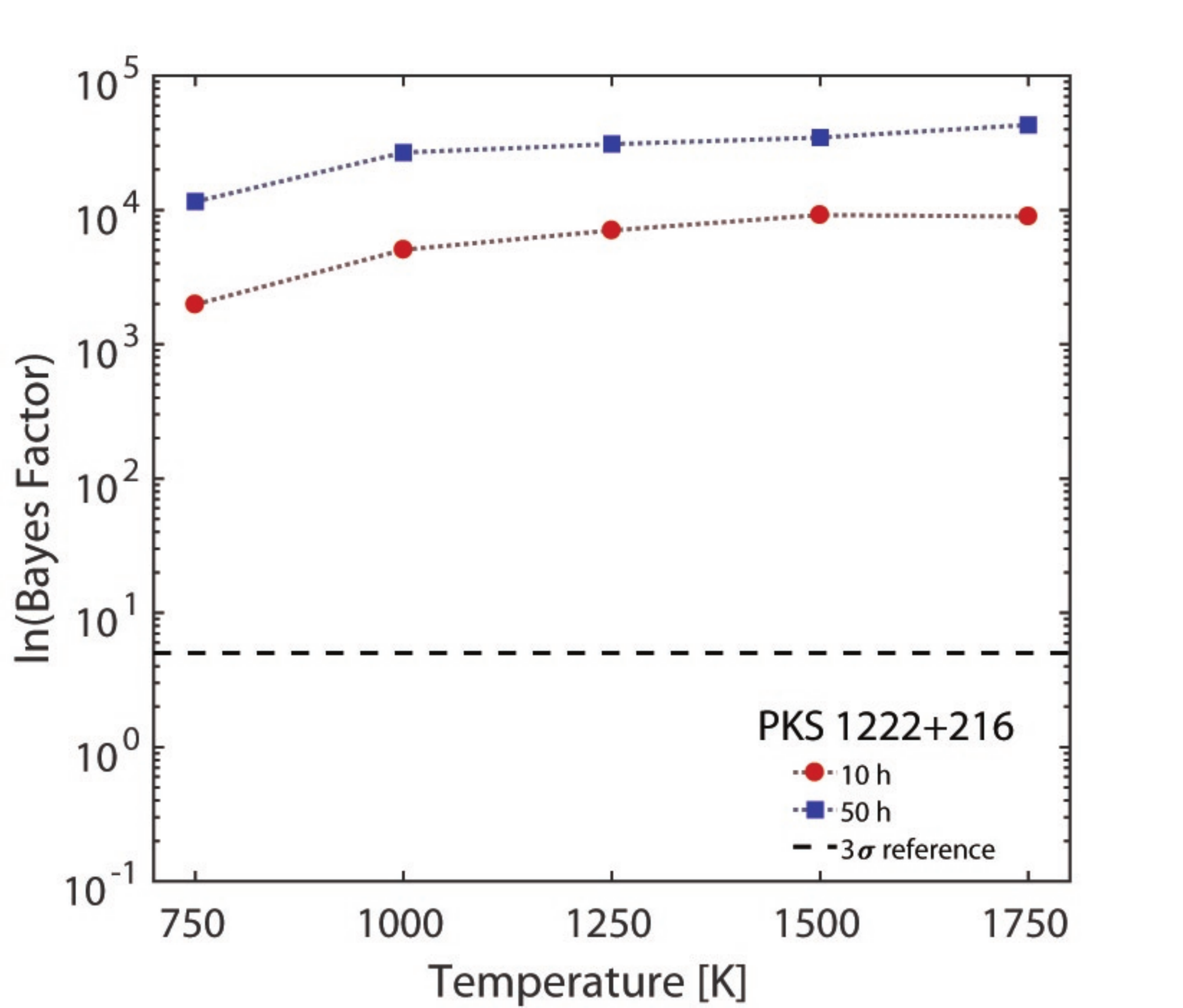}\includegraphics[width=.33\textwidth]{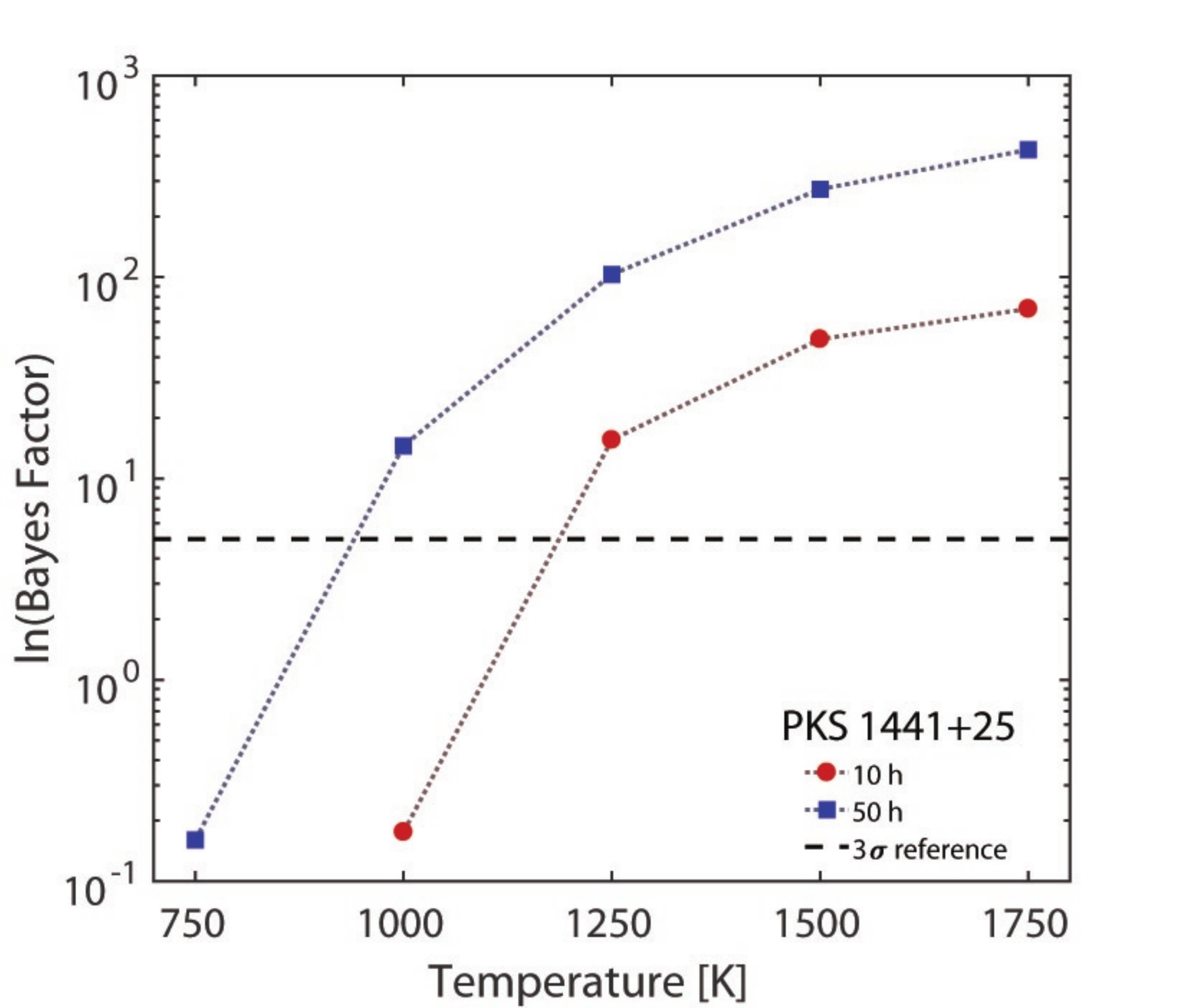}
\end{center}
\caption{\label{bayesF} 
Bayes factors comparing the goodness of the super exponential cut-off power law (SPL) model with respect to the power law (PL) one for PKS 1510-089 (left panel), PKS 1222+216 (central panel) and  PKS 1441+25 (right panel) as a function of the torus temperature and of the exposure time. The dashed black line represents the value of the Bayes factor above which the SPL model is preferred with respect to the PL one.}
\end{figure*}

As a statistical prescription, a model is strongly preferred with respect to another if the corresponding Bayes factor is bigger than 150~\citep{KassRaftery1995}. Since the SPL model is sensible to to torus temperature, the previous statement can be physically converted to the possibility by the CTA in detecting the torus temperature $T$ if the Bayes factor of a particular configuration of the model ($T$, $f_{\rm cov}$) and observational exposure is bigger than 150. In Fig.~\ref{bayesF} we report the Bayes factor comparing the goodness of the SPL model with respect to the PL one for all considered FSRQs as a function of the temperature and of the exposure time.
From Fig.~\ref{bayesF} we observe that for closer sources (PKS 1510-089 and PKS 1222+216) all considered temperatures in the range $750 \, {\rm K}-1750 \, {\rm K}$ are expected to be determined with future CTA data both in the case of 10 h and 50 h of exposure, while for faraway sources (PKS 1441+25) the minimal determinable temperature depends on the exposure time. In particular, we can see that for PKS 1441+25 the CTA is able to determine torus temperature if $T \gtrsim 1200 \, \rm K$ in the case of 10 h of exposure and if $T \gtrsim 900 \, \rm K$ in the case of 50 h of exposure. As a general result, we infer that the best sources to study torus influence are the closest ones since FSRQs at high redshift are deeply affected by EBL absorption for observed energy above $\sim 100 \, \rm GeV$ where the influence of the torus starts to be significant. In addition, we note that, because of cosmic expansion the observed energy, where the torus effects are important, is decreased by a factor $1+z$ but this fact is not enough to avoid the EBL influence. Obviously, by increasing the exposure time we are able to determine lower and lower torus temperatures.

\begin{figure}       
\begin{center}
\includegraphics[width=.45\textwidth]{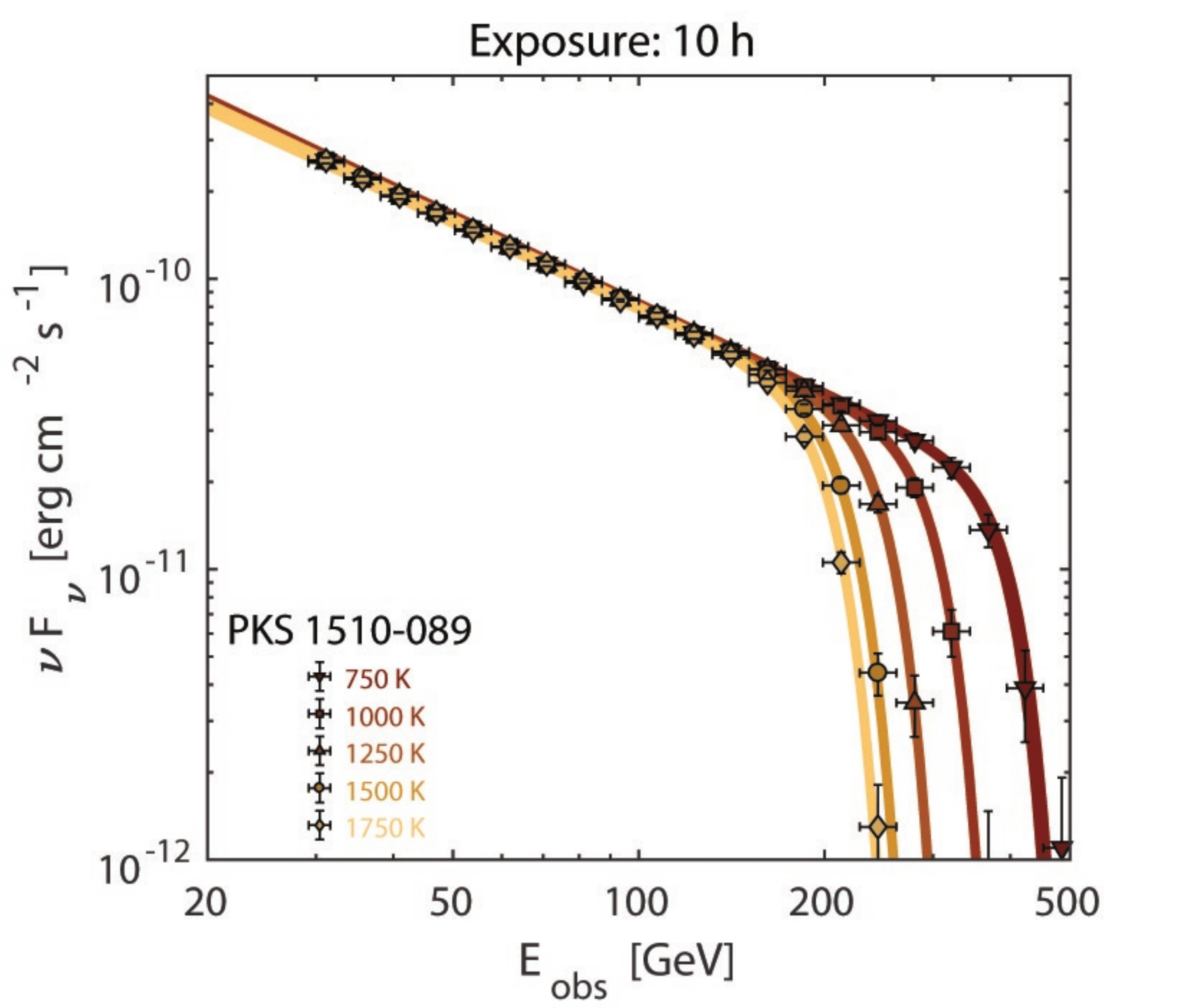}
\includegraphics[width=.45\textwidth]{1510-50h.pdf}
\end{center}
\caption{\label{bins1510} 
EBL-deabsorbed spectrum energy bins for different values of the torus temperature and corresponding fitting functions for PKS 1510-089 in the case of 10 h (upper panel) and 50 h (lower panel) of observation.}
\end{figure}

\begin{figure}       
\begin{center}
\includegraphics[width=.45\textwidth]{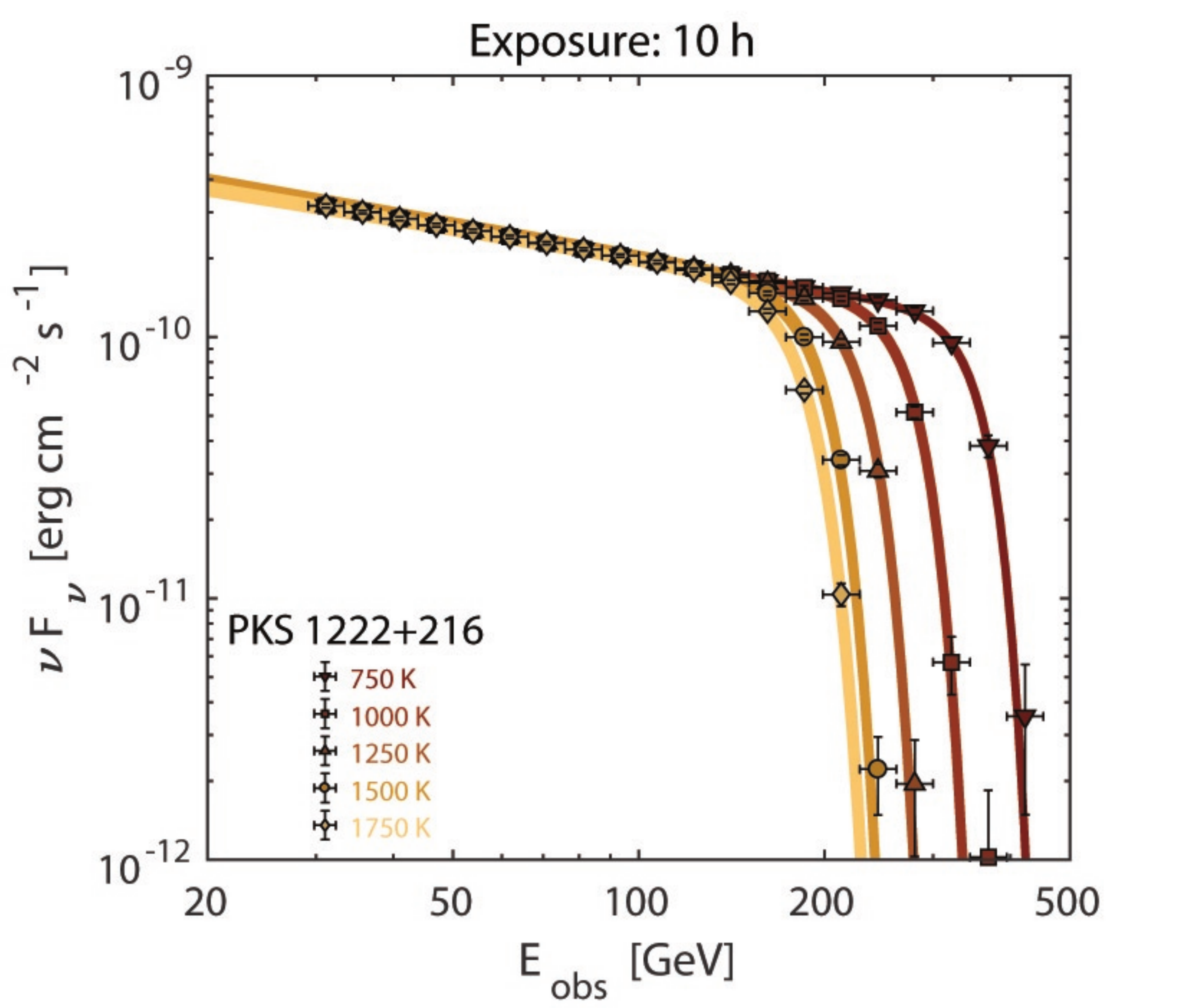}
\includegraphics[width=.45\textwidth]{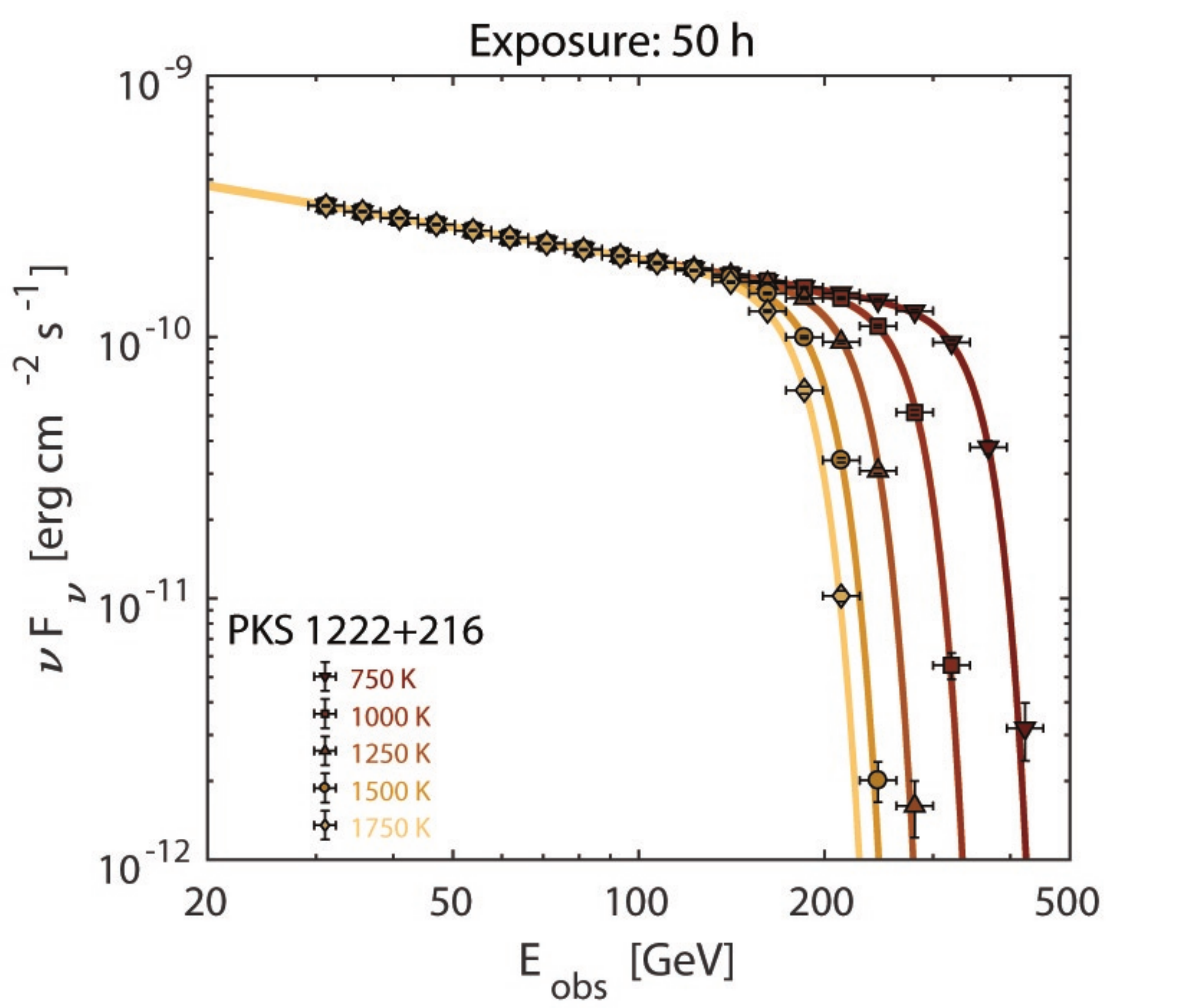}
\end{center}
\caption{\label{bins1222} 
Same as Fig.~\ref{bins1510} but for PKS 1222+216.}
\end{figure}

\begin{figure}       
\begin{center}
\includegraphics[width=.45\textwidth]{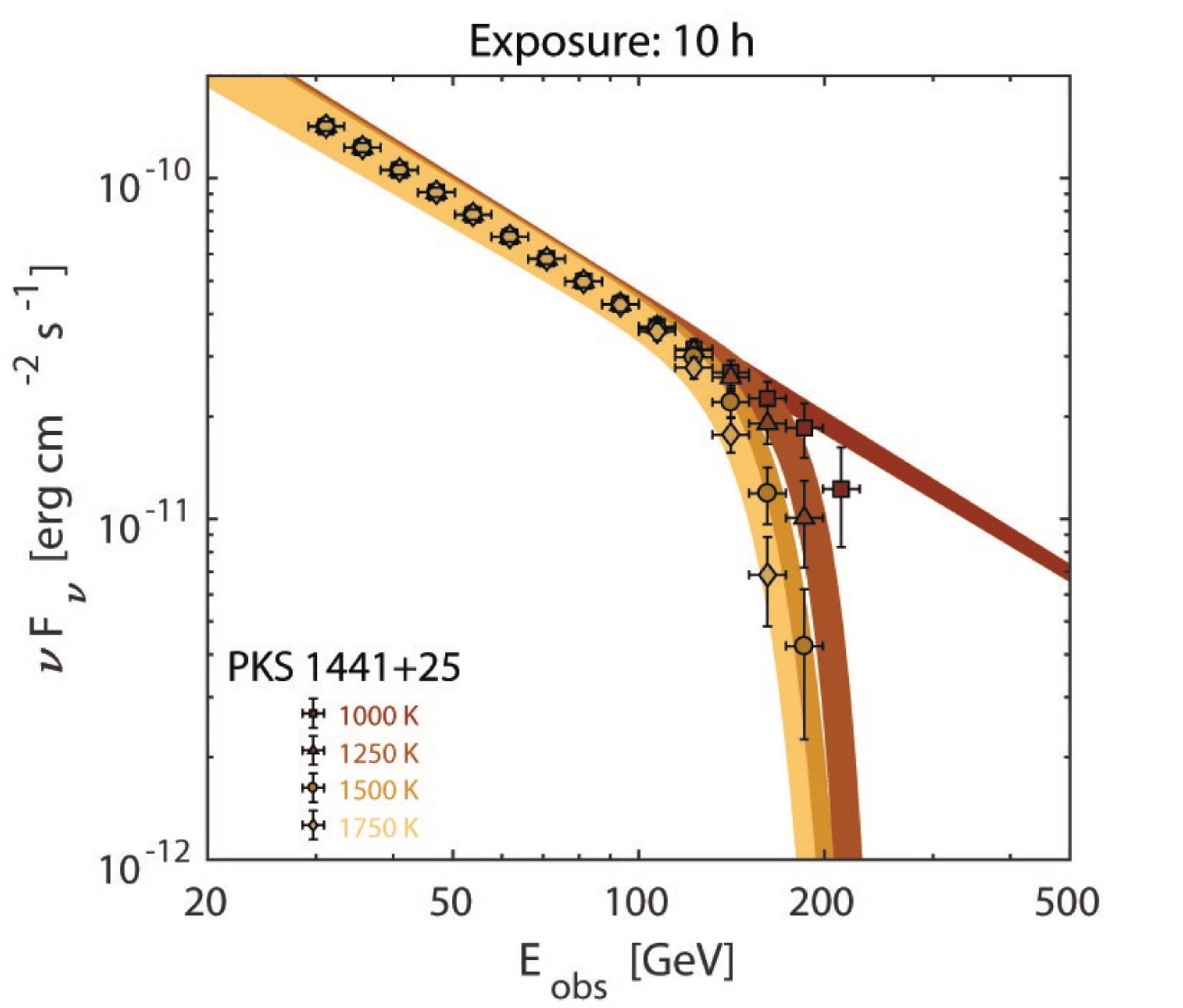}
\includegraphics[width=.45\textwidth]{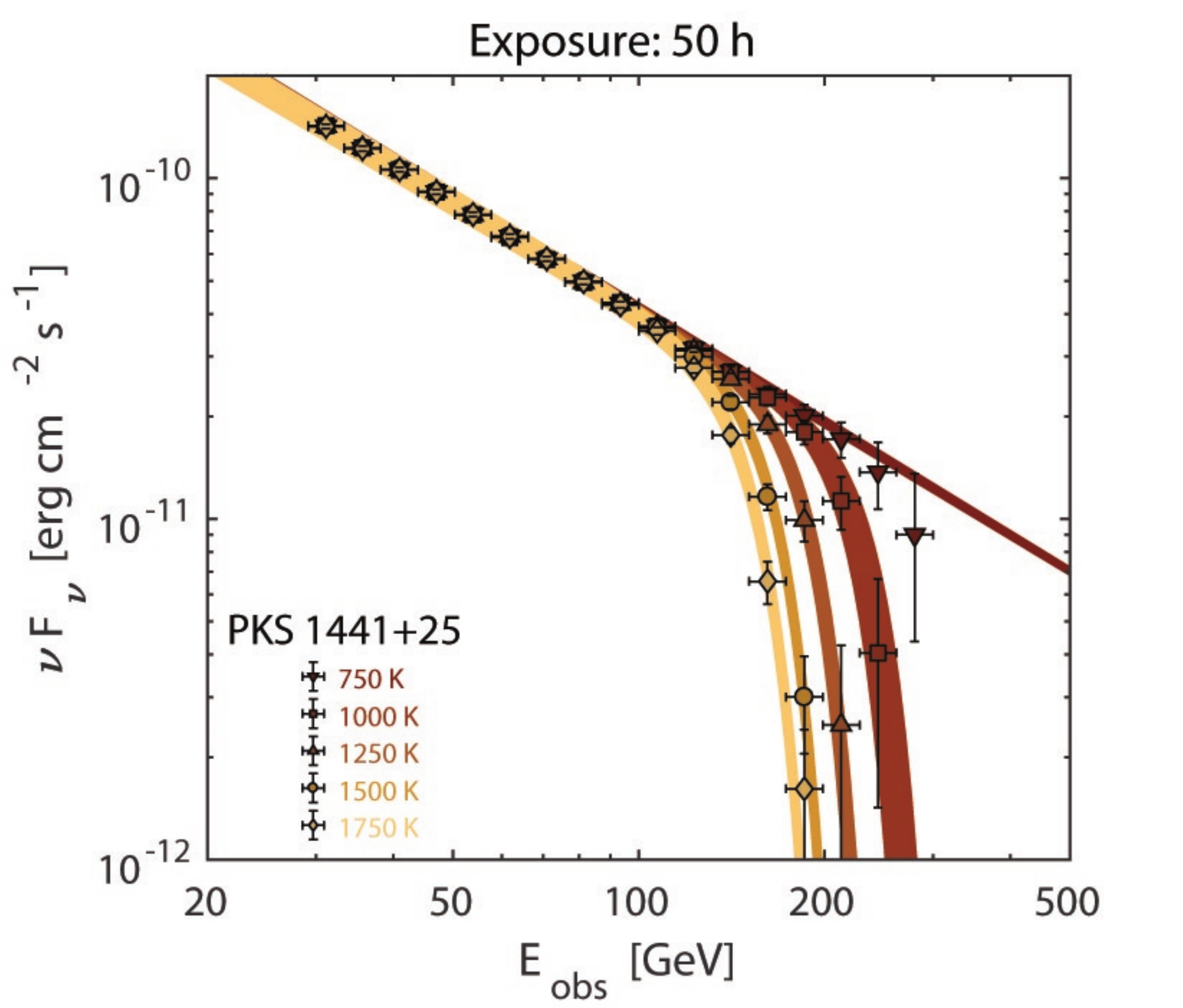}
\end{center}
\caption{\label{bins1441} 
Same as Fig.~\ref{bins1510} but for PKS 1441+25.}
\end{figure}

In Figs.~\ref{bins1510},~\ref{bins1222} and~\ref{bins1441} we report the EBL-deabsorbed spectrum energy bins and the corresponding statistically preferred fitting models for all considered FSRQs and for all considered temperatures in the range $750 \, {\rm K}-1750 \, {\rm K}$ and in both the case of 10 h and 50 h of exposure.
The different thickness of the fitting curves represents the relative errors of the fit. As expected, the statistical error decreases by increasing the exposure time (although one should also take care of the systematic uncertainties) as it is possible to observe by comparing in all the Figs.~\ref{bins1510},~\ref{bins1222} and~\ref{bins1441} the top panels referred to 10 h of exposure with respect to the lower panels corresponding to 50 h of exposure.

We observe that the phenomenological SPL model is very efficient in describing the torus influence for all considered temperatures in the range $750 \, {\rm K}-1750 \, {\rm K}$. As a rule of thumb, the SPL cut-off energy $E_c$ and the torus temperature $T$ are linked by a strict relationship represented by the phenomenological equation
\begin{equation}
\label{TvsEc}
T=1000 \left( \frac{E_c(1+z)}{430 \, {\rm GeV}} \right)^{-1.3} \, \rm K~,
\end{equation}
that links $T$ to the parameter $E_c$  -- which is deducible by observational data -- and to the source redshift $z$. As mentioned above, an alternative -- which will be effective when new data will be available -- is to directly use the model of the optical depth developed in Sect. 2 that allows us to derive directly the torus temperature.
 

At this point, as briefly mentioned in Sect. 3, we want to stress that the EBL-deabsorbed spectrum could deviate from a pure power law because of a curvature of the intrinsic spectrum and/or because of the torus absorption. As studied in~\citet{curvPap} an electron distribution $n_e$ described by a power law with an exponential cut-off 
\begin{equation}
\label{n-ele}
n_e(\gamma_e)=n_{e,0}\,\gamma_e^{-p}e^{-\frac{\gamma_e}{\gamma_c}}~,
\end{equation}
where $\gamma_e$ is the electron Lorentz factor, $\gamma_c$ is the cutoff Lorentz factor, $p$ is the spectral index and $n_{e,0}$ is a normalization constant, produces in the Thomson regime an observed photon SED $\propto {\rm exp}[-(E_{\rm obs}/E_c)^{0.5}]$ which can be fitted by Eq.~(\ref{SPLfit}) where now we call the parameter $\alpha$ for more clarity $\alpha_{\rm intrinsic}=0.5$. The curvature produced by the Klein-Nishina cross section is more involved, but, since the transition between the Thomson and the Klein-Nishima regime is not sharp, we expect that Klein-Nishima effects start to be important at energies well above the energy range where the torus influence is dominant.

\begin{figure}       
\begin{center}
\includegraphics[width=.45\textwidth]{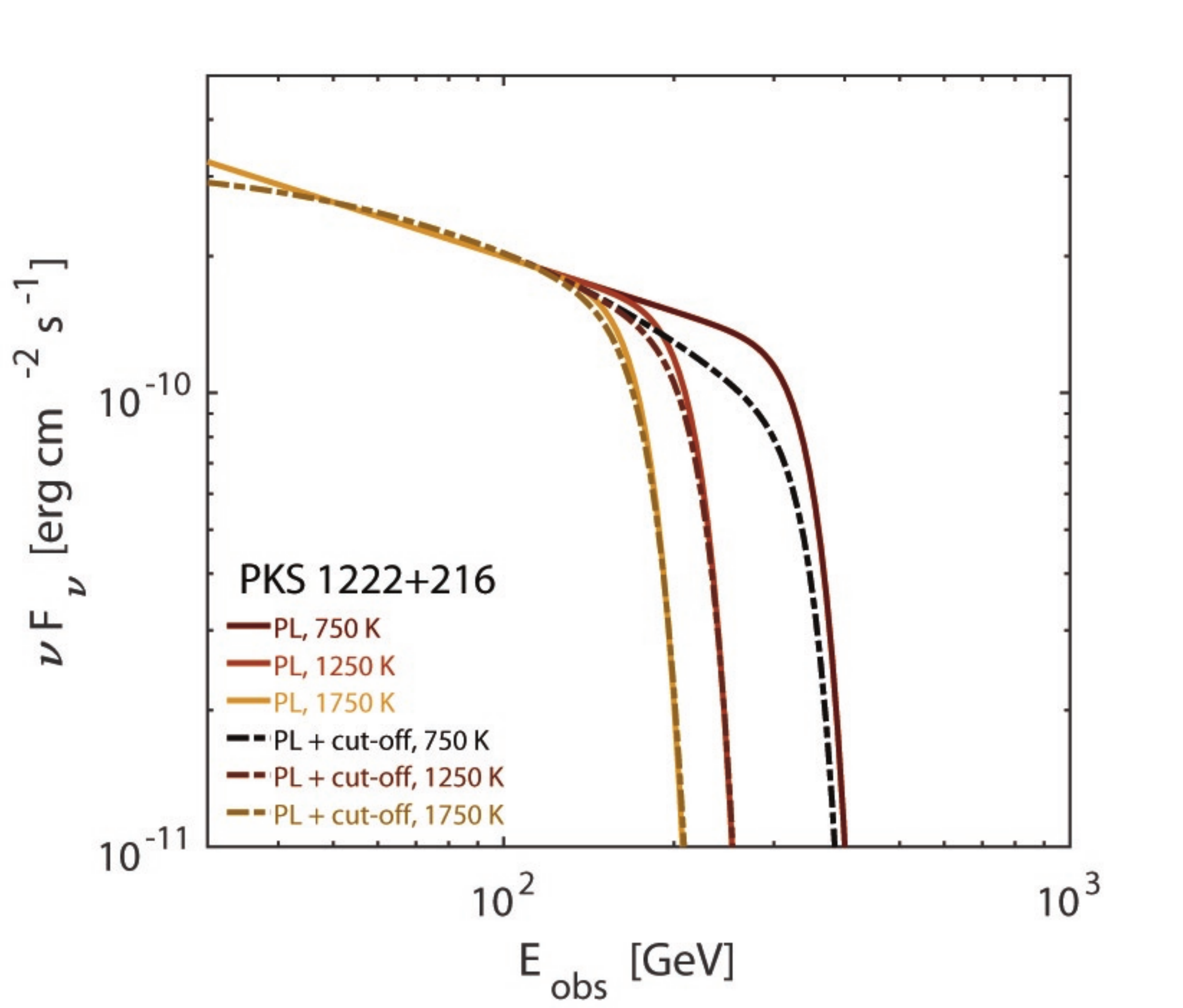}
\end{center}
\caption{\label{curvature} 
EBL-deabsorbed SED of PKS 1222+216 with an intrinsic pure power law (PL, represented by solid lines) spectrum and with additional intrinsic curvature (PL + cut-off, represented by dot-dashed lines). We have fixed $f_{\rm cov}=0.6$. The different  colors refer to different torus temperatures.}
\end{figure}

In Fig.~\ref{curvature} we plot the EBL-deabsorbed spectrum of PKS 1222+216 and we show the different possibilities one can come across. In particular, with the cut-off due to the intrinsic curvature (we name it intrinsic cut-off) fixed at $\sim 50 \, \rm GeV$, we let the value of the torus temperature $T$ vary in the range $750 \, {\rm  K}-1750 \, {\rm K}$ so as to get different values for the cut-off induced by the torus absorption (we name it torus cut-off). For $T=1750 \, \rm K$ the spectrum associated to an intrinsic pure power law is indistinguishable from the one resulting from a power law plus intrinsic curvature at high energies. Nevertheless, the hardness of the curvature associated to the torus (measured by the parameter $\alpha$ in Eq.~(\ref{SPLfit}) -- we call it $\alpha_{\rm torus}$ for more clarity here) appears much bigger than the intrinsic one where $\alpha_{\rm intrinsic}=0.5$~\citep{curvPap} allowing us in principle to identify the torus contribution. For  $T=750 \, \rm K$ we observe that the spectrum induced by an intrinsic pure power law considerably differs from the one induced by a power law plus intrinsic curvature. The identification of the origin of the curvature depends on the possibility to have data also at the energies where the torus absorption influence is important and on the possibility to constrain the $\alpha$ parameter. For $T=1250 \, \rm K$ we have an intermediate situation. 

From our statistical analysis we infer that the torus influence is accurately described by a SPL model of Eq. (\ref{SPLfit}) with $\alpha_{\rm torus} > 8$ and small relative error $<10\%$ (see Fig.~\ref{cornerPlot}). As a result, we can infer that, since the eventual additional intrinsic curvature to the spectrum possesses $\alpha_{\rm intrinsic} = 0.5$ in Eq. (\ref{SPLfit}), it can be distinguished from the one associated to the torus absorption ($\alpha_{\rm torus} >8$).  Curvature associated to the Klein-Nishina cross section can be equally well distinguished since we expect $\alpha_{\rm intrinsic} \simeq 1$ at most in the extreme Klein-Nishina regime.

\section{Discussion and conclusion}
In this paper we have studied the importance of the torus absorption in FSRQs. We use observational data of FSRQs in flaring state since they have a more powerful flux, thus allowing us to study more accurate spectra with better statistics. In addition, we want to stress that a survey to observe flaring FSRQs is foreseen by CTA Key Science Programs~\citep{CTAscience}. We have placed the emission region beyond the BLR in order to avoid BLR absorption to explain the detection of photons coming from FSRQs with energies above $20 \, \rm GeV$ without invoking physics beyond the Standard Model of particle physics~\citep{costamante18}. Thus, we have modeled the torus emission with a black body, so that the torus is characterized by its temperature $T$ and its geometry resumed by the covering factor $f_{\rm cov}$ that measures how much the central supermassive BH is covered by the torus. We have built different observed FSRQ spectra by varying torus parameters and we have inferred that a degeneracy between $T$ and $f_{\rm cov}$ exists. The emission position $r_{\rm em}$ is another parameter that can complicate the situation since it influences the torus opacity, as it is correlated with $f_{\rm cov}$. Thus, we infer that future observational campaigns will be of paramount importance to constrain better $f_{\rm cov}$ and $r_{\rm em}$ in order to determine accurately the torus temperature.

Thus, by fixing $f_{\rm cov}$ to its typical value $f_{\rm cov}=0.6$~\citep{fcov} we have produced different spectrum models for all considered FSRQs (PKS 1510-089, PKS 1222+216 and  PKS 1441+25) by varying the torus temperature in the range $750 \, {\rm K}-1750 \, {\rm K}$. Then, we have produced the spectrum energy bins for all the sources and studied if the CTA will be able to infer torus properties from future observational data. Therefore, we have used two functions to fit the data, a power law (PL), unable to consider the torus influence, and a super exponential cut-off power law (SPL) whose cut-off energy is stricty linked to the torus temperature. We have performed a statistical analysis in order to understand which is the preferred model to fit the data. By studying the Bayes factor of the SPL model with respect to the PL model we have inferred that the SPL model is always preferred for the torus temperatures under consideration when sources at low redshifts ($z \lesssim 0.5$) are considered. On the contrary, if sources at higher redshifts are taken into account, we infer that we obtain a lower limit for the temperature below which the CTA is unable to determine the torus temperature. The reason for this behaviour is that at high redshifts the EBL is so intense that the flux at the energies where torus influence is important gets strongly decreased and especially for low torus temperatures since torus influence happens at higher energies where the EBL is stronger. Obviously, if we increase the exposure time, such a lower limit of detectable torus temperature gets decreased.

Finally, we have deduced a phenomenological equation linking the fitting parameter of the SPL model $E_c$ to the torus temperature $T$, so that fitting future observational data with a SPL model can give us directly information about $T$. If other torus parameters are well constrained ($f_{\rm cov}$, $r_{\rm em}$), a precise detection of $E_c$ would imply a good accuracy in estimating the torus temperature.

We want to state that in this paper we assume an intrinsic pure power law spectrum so that the only curvature on the EBL-deabsorbed spectrum is that associated to the torus absorption. We want to stress that even if other curvatures from intrinsic origin (on-set of the Klein-Nishina regime above ${\cal O}(100 \, \rm GeV)$ and/or due to modifications in the electron spectrum) were present, they should be easily distinguishable from that coming from the torus absorption, since any intrinsic curvature is expected to be much more moderate than the abrupt cut-off induced by torus absorption.


As we have previously stated, placing the photon emission region beyond the BLR is the only possibility in order to avoid BLR absorption if we consider conventional physics. Yet, the existence of axion-like particles (ALPs) and consequently photon-ALP oscillations in the presence of an external magnetic field (see e.g.~\citealt{dgrGen,grSM} and references therein) has many consequences in VHE astrophysics (see e.g.~\citealt{gRew,gtl2019} and references therein). In FSRQs photon-ALP interaction could allow us to place the emission region within the BLR, as in classical AGN leptonic models~\citep{Maraschi92,Sikora94,ssc1}. In fact, as shown in~\citet{trgb2012} photons may convert into ALPs very closely to the emission region inside the jet magnetic field, ALPs are not absorbed since they do not interact with BLR photons and then, ALPs can be converted back to photons outside the BLR in the magnetic field of the jet~\citep{trg}, in the extragalactic magnetic field~\citep{grExt}, in the Milky Way magnetic field or in all these magnetic fields~\citep{gtre}. We want to stress that photon-ALP conversion inside FSRQs may modify our findinds in a sizable way.

\section*{Acknowledgments}
GG and FT acknowledge contribution from the grant INAF CTA--SKA, ``Probing particle acceleration and $\gamma$-ray propagation with CTA and its precursors'' and the INAF Main Stream project ``High-energy extragalactic astrophysics: toward the Cherenkov Telescope Array''.

This paper has gone through an internal review by the CTA Consortium.


\label{lastpage}

\end{document}